\documentclass[a4paper,oneside,11pt,DIV=11]{scrartcl}

\addtokomafont{disposition}{\rmfamily}
\DeclareOldFontCommand{\rm}{\normalfont\rmfamily}{\mathrm}

\usepackage{hyperref}
\usepackage[affil-it]{authblk}
\usepackage{amsmath}
\usepackage{amssymb}
\usepackage{amsthm}
\usepackage{stmaryrd}
\usepackage{xspace}
\usepackage{refcount}
\usepackage[colorinlistoftodos]{todonotes}
\usepackage{enumerate}
\usepackage{accents}

\newtheorem{lemma}{Lemma}
\newtheorem{proposition}{Proposition}
\newtheorem{theorem}{Theorem}
\newtheorem{corollary}{Corollary}
\newtheorem{example}{Example}

\theoremstyle{plain}
\newtheorem{definition}{Definition}

\newtheorem{cus}{Lemma}
\newtheorem{prm}{Lemma}

\newcommand{\intp}[1]{\llbracket #1 \rrbracket}
\newcommand{\FO}{\textnormal{FO}}
\newcommand{\DA}{\mathbf{DA}}

\newcommand{\Mvar}[1]{\mathbf{Si}_{#1}}
\newcommand{\Th}{\textsuperscript{th}\xspace}
\newcommand{\faktor}[2]{{#1 / \! #2}}

\newcommand{\alp}{{\mathrm{alph}}}
\newcommand{\im}{{\mathrm{im}}}

\newcommand{\greenfont}[1] {\ensuremath{\mathcal{#1}}}

\newcommand{\greenR} {\greenfont{R}\xspace}
\newcommand{\Req}    {\mathrel{\greenR}}
\newcommand{\Rleq}   {\mathrel{\leq_\greenR}}
\newcommand{\Rl}     {\mathrel{<_\greenR}}

\newcommand{\greenL} {\greenfont{L}\xspace}
\newcommand{\Leq}    {\mathrel{\greenL}}
\newcommand{\Lleq}   {\mathrel{\leq_\greenL}}
\newcommand{\Ll}     {\mathrel{<_\greenL}}

\newcommand{\greenJ} {\greenfont{J}\xspace}
\newcommand{\Jeq}    {\mathrel{\greenJ}}
\newcommand{\Jleq}   {\mathrel{\leq_\greenJ}}

\title{Nesting Negations in FO$^2$ \\ over Infinite Words}
\author{Viktor Henriksson\hspace*{1pt}$^1$ and Manfred Kuf\-leitner\hspace*{2pt}$^2$%
	\affil{Loughborough University, Loughborough, UK}}

\date{\normalsize$^1$ Loughborough University, Loughborough, UK \\
\texttt{B.V.D.Henriksson@lboro.ac.uk} \\[3mm]
$^2$ University of Stuttgart, Stuttgart, Germany \\
\texttt{kufleitner@fmi.uni-stuttgart.de}}

\begin{document}

\maketitle

\begin{abstract}
  \noindent
  \textbf{Abstract.}\;
We consider two-variable first-order logic $\FO^2$ over infinite words. Restricting the number of nested negations defines an infinite hierarchy; its levels are often called the half-levels of the $\FO^2$ quantifier alternation hierarchy. For every level of this hierarchy, we give an effective characterization. For the lower levels, this characterization is a combination of an algebraic and a topological property. For the higher levels, algebraic properties turn out to be sufficient. Within two-variable first-order logic, each algebraic property is a single ordered identity of omega-terms. The topological properties are the same as for the lower half-levels of the quantifier alternation hierarchy without the two-variable restriction (i.e., the Cantor topology and the alphabetic topology).
  
Our result generalizes the corresponding result for finite words. The proof uses novel techniques and is based on a refinement of Mal'cev products for ordered monoids.
\end{abstract}

\section{Introduction}

Connections between regular languages, finite monoids and logic have a long tradition in theoretical computer science. Eilenberg's Variety Theorem gives a one-to-one correspondence between varieties of formal languages and varieties of finite monoids~\cite{eilenberg1974}. While the difficult direction of this result usually is not used when establishing such correspondences for concrete varieties, it points towards a suitable algebraic counterpart for recognizing a given language class. For instance, Sch\"utzenberger showed that a language is star-free if and only if it is recognized by a finite aperiodic monoid~\cite{shutzenberger1976sf}. Every language is recognized by a unique minimal monoid, the syntactic monoid of the language~\cite{RabinScott1959}. Hence, by Sch\"utzenberger's Theorem one can decide whether a given regular language is star-free: one computes its syntactic monoid and checks whether it is aperiodic.

McNaughton and Papert showed that a language is definable in first-order logic if and only if it is star-free~\cite{mcnaughtonpapert1971}. Together with Sch\"utzenberger's Theorem, this yields decidability of the question whether a given regular language is definable in first-order logic. In addition, the connection to aperiodic monoids leads to new insights into the first-order definable properties; e.g., it immediately shows that one cannot define any modular counting within first-order logic over finite words.

Eilenberg's Variety Theorem was generalized in several directions. Pin considered language classes which might not be closed under complement and showed that finite ordered monoids are a suitable algebraic counterpart~\cite{pin1995rm}. There exist several approaches for extending Eilenberg's Theorem to infinite words. The most principled generalizations are in terms of Wilke-algebras~\cite{wilke1993ijac} and $\omega$-semigroups; see~\cite[Chapter~II]{PerrinPin2004}. During the last few years, these connections have been further generalized to more abstract settings; see e.g.~\cite{Bojanczyk2015dlt,UrbatEtAl2017mfcs}. On the other hand, for full first-order logic over infinite words, a slightly simpler approach based on the recognition by finite monoids and Arnold's syntactic congruence~\cite{arnold1985tcs} is sufficient; see e.g.~\cite{DiekertGastin2008}. Intuitively, simpler algebraic recognizers are often more desirable because they more easily reveal properties of the corresponding languages. Over infinite words, however, there are many interesting language classes (such as fragments of first-order logic) which cannot be characterized in terms of Arnold's congruence; see e.g.~\cite[Chapter~VIII]{PerrinPin2004}. On the other hand, the combination of such algebraic properties with topology can provide characterizations of various language classes; see~e.g~\cite{diekertkufleitner2011tocs,KallasEtAl2011stacs,KufleitnerWalter2018tocs}. Usually, the topological conditions restrict the \emph{behavior at infinity}.

The use of logical fragments for defining regular languages dates back to the early 1960s, when B\"uchi, Elgot and Trakhtenbrot independently showed that a language over finite words is regular if and only if it is definable in monadic second-order logic~\cite{buchi1960zmlgm,elgot1961tams,trakhtenbrot1961spd}. This result was extended to infinite words by 
B\"uchi~\cite{buchi1960zmlgm,Buchi1962}.
 Other classic results over finite words concern first-order logic~\cite{mcnaughtonpapert1971,shutzenberger1976sf} and the quantifier alternation hierarchy in first-order logic~\cite{thomas1982jcss}. Later,  
in 1998, Th\'erien and Wilke~\cite{TherienWilke1998stoc} investigated two-variable first-order logic $\FO^2$ over finite words; by definition, the set of variables in $\FO^2$ is restricted to, say, $x$ and $y$ which can arbitrarily be reused. Among other results, they established a correspondence between $\FO^2$ and the variety~$\DA$ in the spirit of Eilenberg's Variety Theorem.

Over finite words, the fragment $\FO^2$ is contained within the fragment $\Sigma_2$ of the quantifier alternation hierarchy~\cite{TherienWilke1998stoc}. However, one can also consider the quantifier alternation hierarchy within $\FO^2$; here -- due to the restricted number of variables -- one needs to rely on parse trees rather than equivalent formulae in prenex normal form. The $\FO^2$ (quantifier) alternation hierarchy was first considered by Weis and Immerman \cite{weisimmerman2009lmcs}; algebraic characterizations were given by Weil and the second author~\cite{kufleitnerweil2012csl}  and independently by Krebs and Straubing~\cite{KrebsStraubing2017tocl}. The proof in~\cite{kufleitnerweil2012csl} uses Mal'cev products with definite and reverse-definite semigroups. These operations can be expressed by congruences $\sim_{\mathbf{D}}$ and $\sim_{\mathbf{K}}$; see~\cite{krt68arbib8}. Krebs and Straubing's proof uses block products. This technique was adapted to ordered monoids and led to a characterization of the half-levels of the $\FO^2$ quantifier alternation hierarchy (denoted by~$\Sigma^2_m$) over finite words~\cite{FleischerKL17tocs}.

In this paper, we characterize the half levels of the $\FO^2$ alternation hierarchy over finite and infinite words using a combination of algebra and topology. Block products (both in their ordered and unordered version) over finite words satisfy some left-right duality which does not hold for infinite words. Thus, block products seem to be a less convenient tool for generalizing a proof from finite to infinite words. Our approach is based on a new relation $\preceq_{\mathbf{KD}}$. This partial order is inspired by $\sim_{\mathbf{K}}$ and $\sim_{\mathbf{D}}$, but adapted for dealing with ordered monoids. We introduce a chain of varieties $\Mvar{m}$ which can be built inductively using the relation $\preceq_{\mathbf{KD}}$. We show that these varieties correspond to the monoids in $\DA$ which satisfy the identities $U_m \leq V_m$ introduced in \cite{FleischerKL17tocs}; in particular, our result yields a new proof for this characterization over finite words.

Our main result is an effective characterization of $\Sigma^2_m$ for each $m \geq 1$. An $\omega$-language (i.e., a language over infinite words) is definable in $\Sigma^2_1$ if and only if it is both open in the Cantor topology (cf.~\cite[Chapter III.3]{PerrinPin2004}) and recognized by a monoid in $\Mvar{1} = \intp{1 \leq z}$. An $\omega$-language is definable in~$\Sigma^2_2$ if and only if it is both open in the alphabetic topology (cf.~\cite{diekertkufleitner2011tocs}) and recognized by a monoid in $\Mvar{2}$. For $m \geq 3$, an $\omega$-language is definable in $\Sigma^2_m$ if and only if it is recognized by a monoid in $\Mvar{m}$.

While the characterization of $\Sigma^2_1$ for $m=1$ is straightforward (and considered only for completeness), the proofs for $m\geq 2$ are more involved. We use Ehrenfeucht-Fra\"iss\'e games to show that definability in $\Sigma^2_m$ yields recognizability in $\Mvar{m}$. The bigger challenge is the proof of the other direction. We consider two orders on the set of infinite words: One order is induced by the logical fragment and the other order is given by the algebraic requirement. By induction on $m$, we show that the logical order refines the algebraic order. However, this step only holds for words with the same letters occurring infinitely often. Since the set of letters occurring infinitely often can be defined in $\Sigma^2_m$ only if $m \geq 3$, the case $m=2$ requires some further considerations and is technically more involved. The restriction of being open in the alphabetic topology compensates for this lack of expressivity of $\Sigma^2_2$.

\section{Preliminaries}

Let $A$ be an alphabet. The set of all finite words over $A$ is $A^*$, and $A^\omega$ is the set of all infinite words. The set of all words is $A^\infty = A^* \cup A^\omega$. A word $a_1 \cdots a_n \in A^*$ is a (scattered) \emph{subword} of $\alpha \in A^\infty$ if there exist words $v_i \in A^*$ and $\alpha' \in A^\infty$ such that $\alpha = v_1 a_1 v_2 \cdots v_n a_n \alpha'$.
For a finite word $u$, its infinite iteration is $u^\omega = u u u \cdots$. The empty words is $\varepsilon$ and it satisfies $\varepsilon^\omega = \varepsilon$. For a set $U\subseteq A^*$, the language $U^{\omega}$ contains all words $u_1 u_2 u_3 \cdots$ with $u_i \in U$.
For $\alpha \in A^\infty$, let $\alp(\alpha)$ be the set of letters occurring in $\alpha$. The \emph{imaginary} alphabet $\im(\alpha)$ is the set of letters occurring infinitely often in $\alpha$.

\subsection{Monoids and Varieties}

A binary relation $\preceq$ on a monoid $M$ is \emph{stable} if $s \preceq t$ implies $xsy \preceq xty$ for all $s,t,x,y \in M$. An \emph{ordered monoid} is a monoid together with a stable partial order $\leq$ on its elements. Every monoid can be equipped with a stable partial order by using the equality. A  monoid homomorphism $\mu: M \to N$ is \emph{monotone} if $x \leq y$ implies $\mu(x) \leq \mu(y)$ for all $x,y \in M$. Unless stated otherwise, we assume that all homomorphisms are monotone. An ordered monoid $N$ \emph{divides} an ordered monoid $M$ if there exists a submonoid $N' \subseteq M$ and a surjective monotone homomorphism $N' \to N$. It is also possible to define division without the assumption that the surjective homomorphism is monotone; in this case, we speak of \emph{unordered division} even though the monoids are ordered. A \emph{preorder} $\preceq$ on $M$ is a reflexive and transitive binary relation. An element $e \in M$ is \emph{idempotent} if $e^2 = e$.
For every finite monoid $M$, there exists an integer $\omega_M \geq 1$ such that all $s \in M$ satisfy $s^{\omega_M}s^{\omega_M} = s^{\omega_M}$, i.e., $s^{\omega_M}$ is idempotent. When $M$ is clear from context, we may also denote the number by $\omega$.

A \emph{positive variety} is a class of finite ordered monoids closed under finite products and (monotone) division. A \emph{variety} is defined similarly, using unordered division. We can define varieties and positive varieties using identities of $\omega$-terms. Let $\Omega$ be a set of variables. The set $T(\Omega)$ of \emph{$\omega$-terms} over $\Omega$ is defined as follows: $\Omega \cup \left\{ 1 \right\} \subseteq T(\Omega)$.
Furthermore, if $u,v \in T(\Omega)$, then the formal product $uv$ and the formal $\omega$-power $u^{\omega}$ are also in $T(\Omega)$. An $\omega$-identity has the form $u \leq v$ or $u=v$ for $\omega$-terms $u,v \in T(\Omega)$. Given a finite monoid $M$, every interpretation $I: \Omega \to M$ can be extended to a map $I: T(\Omega) \to M$ by $I(1) = 1$, $I(uv) = I(u)I(v)$ and $I(u^{\omega}) = I(t)^{\omega_M}$. A monoid $M$ satisfies an $\omega$-identity $u \leq v$ if $I(u) \leq I(v)$ for all interpretations $I : \Omega \to M$; satisfiability of $u=v$ is defined similarly. The class of finite ordered monoids satisfying $u \leq v$ is denoted by $\intp{u \leq v}$; similarly, $\intp{u = v}$ is the class of finite ordered monoids satisfying $u = v$.
Let 
\begin{itemize}
\item $\mathbf{A} = \intp{x^{\omega}x = x^{\omega}}$,
\item $\DA = \intp{(xyz)^{\omega}y(xyz)^{\omega} = (xyz)^{\omega}}$, and 
\item $\mathbf{J}^+ = \intp{1 \leq z}$.
\end{itemize}
We have $\mathbf{J}^+ \subsetneq \DA \subseteq \mathbf{A}$.
The use of $\omega$ in the context of finite monoids is not to be confused with ordinal $\omega$ used in infinite concatenations.

A pair $(s,e) \in M^2$ is \emph{linked} if $se = s$ and $e^2 = e$. Two linked pairs $(s,e)$ and $(t,f)$ are \emph{conjugated}, denoted by $(s,e) \sim (t,f)$, if there exist $x$, $y$ such that $sx = t$, $xy = e$ and $yx = f$. We also consider Green's relations \greenR , \greenL  and \greenJ. For $s,t \in M$, we have
\begin{alignat*}{7}
  s &\Rleq t &&\ \ \Leftrightarrow\ \ & sM &\subseteq tM, &&\qquad\qquad& s &\Req t &&\ \ \Leftrightarrow\ \ & sM &= tM,
  \\ s &\Lleq t &&\ \ \Leftrightarrow\ \ & Ms &\subseteq Mt, &&\qquad& s &\Leq t &&\ \ \Leftrightarrow\ \ & Ms &= Mt,
  \\ s &\Jleq t &&\ \ \Leftrightarrow\ \ & MsM &\subseteq MtM, &&\qquad& s &\Jeq t &&\ \ \Leftrightarrow\ \ & MsM &= MtM.
\end{alignat*}
We write $s \Rl t$ if $s \Rleq t$ but not $s \Req t$, and similarly for $s \Ll t$. Note that $s \Rleq t$ can be equivalently defined as there existing $y \in M$ such that $ty = s$, and symmetrically for $\Lleq$. Correspondingly, $s \Jleq t$ means there exist $x, y \in M$ such that $xty = s$. The following property of $\DA$ is well-known; see e.g.~\cite{diekertgastinkufleitner2008ijfcs} for a proof.

\begin{lemma}\label{lem:DA}
  Let $M \in \DA$ and let $e,a_1,\ldots,a_k\in M$ such that $e$ is idempotent and $e \Jleq a_i$ for all $i$. Then $e a_1 \cdots a_k e = e$.
\end{lemma}

We record the following lemma which is one of the main combinatorial properties of the variety $\DA$. Essentially, it is a consequence of Lemma~\ref{lem:DA}; see e.g.~\cite{kufleitnerweil2012csl}. A typical application is for the homomorphism $M^* \to M$ which evaluates words over the alphabet $M$ by their product in $M$.

\begin{lemma}\label{lem:alphabetLR}
	Let $\mu: A^* \to M \in \DA$ be a homomorphism and let $u,v,w \in A^*$. If $\mu(uv) \Req \mu(u)$ and $\alp(w) \subseteq \alp(v)$, then $\mu(uw) \Req \mu(u)$. Similarly, if $\mu(vu) \Leq \mu(u)$ and $\alp(w) \subseteq \alp(v)$, then $\mu(wu) \Leq \mu(u)$.
\end{lemma}

\subsection{Languages and Recognition}\label{ssc:rec}

A \emph{language} is a subset of $A^\infty$. Let $M$ be a finite monoid and suppose that $\mu: A^* \to M$ is a homomorphism. For $s \in M$, we write $[s]$ to denote $\mu^{-1}(s)$ when $\mu$ is clear from context. For every word $\alpha\in A^\infty$ there exists a linked pair $(s,e)$ such that $\alpha \in [s][e]^\omega$; see e.g.~\cite[Lemma~7.56]{DiekertEtAl2016}. Note however that, in general, such a linked pair is not unique. An ordered monoid $M$ \emph{(strongly) recognizes} a language~$L$ if there exists a homomorphism $\mu : A^* \to M$ such that the following two properties hold:
\begin{itemize}
\item $L = \bigcup \left\{ [s][e]^{\omega} \mid (s,e) \text{ is a linked pair and } [s][e]^{\omega} \cap L \neq \emptyset \right\}$
\item For all $s,t,e \in M$ with $s \leq t$: if $[s][e]^{\omega} \subseteq L$, then $[t][e]^{\omega} \subseteq L$.
\end{itemize}
If the latter property holds, we say that $L$ is \emph{upward closed}.  
A language $L$ is \emph{$\omega$-regular} if there exists a finite monoid $M$ which recognizes $L$. If $M$ recognizes $L$ and $(s,e) \sim (t,f)$, then $[s][e]^{\omega} \subseteq L$ if and only if $[t][f]^{\omega} \subseteq L$.

The \emph{syntactic preorder} $\leq_L$ of a language $L$ is defined by $u \leq_L v$
\begin{equation*}
	xuyz^{\omega} \in L \;\Rightarrow\; xvyz^{\omega} \in L \quad \text{and} \quad x(uy)^{\omega} \in L \;\Rightarrow\; x(vy)^{\omega} \in L.
\end{equation*}
for all $x,y,z \in A^*$. The \emph{syntactic congruence} is defined as $u \equiv_L v$ if $u \leq_L v$ and $v \leq_L u$. The \emph{syntactic monoid} of $L$ is the ordered monoid $\faktor{A^*}{\leq_L}$; its elements are the congruence classes of $\equiv_L$ and the order is induced by $\leq_L$. The natural projection $\mu_L: A^* \to \faktor{A^*}{\leq_L}$ is called the \emph{syntactic homomorphism}. If $L$ is $\omega$-regular, then the syntactic monoid recognizes $L$ via the syntactic homomorphism. Moreover, the syntactic monoid is minimal in the following sense: if $L$ is recognized by a finite monoid $M$, then the syntactic monoid of $L$ divides $M$.

There exist many results regarding the connection between language classes and varieties. For instance, an $\omega$-regular language $L$ is definable in $\FO^2[<]$ if and only if its syntactic monoid is in~$\DA$~\cite{diekertkufleitner2011tocs}.

\begin{example}\label{exa:alp}
  For every alphabet $A$, we can consider the powerset $2^A$. This forms an ordered monoid by using union for composition and set inclusion as the partial order, i.e., we have $B \leq C$ for $B,C \subseteq A$ if $B \subseteq C$. This monoid satisfies $2^A \in \mathbf{J}^+ = \intp{1 \leq z}$. By mapping every word $u$ to its alphabet $\alp(u)$, we define a homomorphism $\alp : A^* \to 2^A$. It can recognize unions of languages of the form $L_{B,C} = \{ \alpha \in A^\infty \mid B \subseteq \alp(\alpha) \text{ and } C = \im(\alpha) \}$ for $C \subseteq B$. 

A very common technique is making a given homomorphism $\mu : A^* \to M$ alphabetic. This means that we consider $\mu' : A^* \to M \times 2^A$ with $\mu'(u) = (\mu(u),\alp(u))$. Note that $\mu'(u) = \mu'(v)$ implies $\alp(u) = \alp(v)$. In particular, $\mu'(u)$ is the neutral element if and only if $u$ is empty. Note that if $M$ is in a variety $\mathbf{V}$ with $\intp{1\leq z} \subseteq \mathbf{V}$, then so is $M \times 2^A$. In other words, we can make homomorphisms alphabetic within a variety $\mathbf{V}$ as soon as $2^A \in \mathbf{V}$.
\end{example}

Next, we introduce a relation $\lesssim$ on linked pairs as an ordered version of conjugacy. In order to not having to switch back and forth between elements and linked pairs, we set $(r,f) \leq (t,f)$ if $r \leq t$.

\begin{definition}
	Let $(s,e)$ and $(t,f)$ be linked pairs over some monoid $M$. We set $(s,e) \lesssim (t,f)$ if there exists a linked pair $(r,f)$ such that  $(s,e) \sim (r,f) \leq (t,f)$.
\end{definition}

This relation is defined to be closely related to language recognition. Suppose that $M$ recognizes $L \subseteq A^{\infty}$. It is easy to see that if $(s,e) \lesssim (t,f)$, then $[s][e]^{\omega} \subseteq L$ implies $[t][f]^{\omega} \subseteq L$.

\begin{lemma}\label{lem:transitivityoflesssim}
	The relation $\lesssim$ is a preorder (i.e., reflexive and transitive).
\end{lemma}

\begin{proof}
	Reflexivity is obvious. For transitivity, consider the chain
	\begin{equation*}
		(s,e) \sim (r_1,g) \leq (r_2,g) \sim (r_3,f) \leq (t,f).
	\end{equation*}
	By definition of conjugacy, there exist $x,y,x',y'$ such that $e = xy$, $g = yx = y'x'$, $f = x'y'$, $r_1 = sx$ and $r_3 = r_2 y'$. Let $r = r_1 y'$. We have $rf = r_1y'x'y' = r_1g y' = r_1y' = r$, thereby showing that $(r,f)$ is a linked pair. Since $r = r_1y' \leq r_2y' = r_3 \leq t$, we see that $(r,f) \leq (t,f)$.  Moreover, $r = r_1 y' = s xy'$, $e = e^2 = xyxy = xy'  x'y$ and $f = f^2 = x'y'x'y' = x'yxy'$. This shows that $(s,e) \sim (r,f)$.
\end{proof}

The following lemma allows us to apply results of $A^\infty$ to finite words.

\begin{lemma}\label{lem:makefinite}
  Let $M \in \mathbf{A}$. If $(s,1) \lesssim (t,1)$, then $s \leq t$.
\end{lemma}

\begin{proof}
  By definition, there exists $r \in M$ with $(s,1) \sim (r,1) \leq (t,1)$. Hence, there exist $x,y \in M$ with $1=xy=yx$ and $sx = r$. Since $M \in \mathbf{A}$, we conclude that $1=xy = x \cdot 1 \cdot y = x\cdot xy \cdot y = x^\omega y^\omega = x x^{\omega} y^\omega = x \cdot 1 = x$ and, hence, $s=r$.
\end{proof}

Given a set $X$, a \emph{topology} on $X$ is a set $\mathcal{T}$ of subsets of $X$ such that $\emptyset, X \in \mathcal{T}$ and $\mathcal{T}$ is closed under arbitrary unions and finite intersections. A subset of $X$ is \emph{open} if it is in $\mathcal{T}$, and \emph{closed} if its complement is in $\mathcal{T}$. A subset $\mathcal{B} \subseteq \mathcal{T}$ is a 
 \emph{base} $\mathcal{B}$ if every set in $\mathcal{T}$ is a union of sets in $\mathcal{B}$.

Let $A$ be an alphabet. Then the \emph{Cantor topology} on $A^{\infty}$ is the topology generated by the basis $\left\{ uA^{\infty} \mid u \in A^* \right\}$. Similarly, the \emph{alphabetic topology} is generated by the basis $\left\{ uB^{\infty} \mid u \in A^* \text{ and } B \subseteq A \right\}$. Both, the $\omega$-regular open sets of the Cantor topology~\cite[Theorem~VI.3.1]{PerrinPin2004} and the alphabetic topology~\cite{diekertkufleitner2011tocs} are decidable.

\subsection{First Order Logic}

We consider fragments of first-order logic $\FO[<]$ using the order predicate $<$. The syntax is given by
\begin{equation*}
	\varphi ::= \top \mid \bot \mid \lambda(x) = a \mid x = y \mid x < y \mid \neg\varphi \mid \varphi \wedge \varphi \mid \varphi \vee \varphi \mid \exists x \varphi.
\end{equation*}
Here, $a \in A$ is a letter of some finite alphabet $A$, and $x$ and $y$ are variables. We define the macros $\forall x \varphi$ to mean $\neg \exists x \neg \varphi$, and $x\leq y$ to mean $(x = y) \vee (x < y)$. We interpret variables as positions of words, where $<$ is the natural order on integers. For $i \in \mathbb{N}$, the predicate $\lambda(i) = a$ is true for a word $u$ if the $i$\Th position of $u$ is $a$. The semantics of Boolean connectives and existential quantifiers is as usual. For a formula $\varphi$ with a free variable $x$ and for a position $i$ of a word~$\alpha$, we say that $\alpha,i \vDash \varphi$ if $\varphi$  with each occurrence of $x$ replaced by $i$ holds on $\alpha$. This generalizes to any number of variables. A formula $\varphi$ with no free is a \emph{sentence}; we write $\alpha \vDash \varphi$ if $\varphi$ is true when interpreted over $\alpha$. This way, $\varphi$ defines the language $L(\varphi) = \left\{ \alpha \in A^\infty \mid \alpha \vDash \varphi \right\}$. A language $L$ is definable in some first order fragment $\mathcal{F}$ if there exists a sentence $\varphi \in \mathcal{F}$ such that $L = L(\varphi)$.

The $m$th level of the \emph{negation nesting} hierarchy of $\FO$ with quantifier depth $n$ is given by the following formulae:
\begin{align*}
	\varphi_{0,0} &::= \top \mid \bot \mid \lambda(x) = a \mid x=y \mid x<y \mid \neg \varphi_{0} \mid \varphi_0 \vee \varphi_0 \mid \varphi_0 \wedge \varphi_0\\
	\varphi_{m,n} &::= \varphi_{m-1,n} \mid \varphi_{m,n-1} \mid \neg \varphi_{m-1,n} \mid \varphi_{m,n} \vee \varphi_{m,n} \mid \varphi_{m,n} \wedge \varphi_{m,n} \mid \exists x \varphi_{m,n-1}
\end{align*}
The fragment $\Sigma_{m,n}$ consists of all formulae $\varphi_{m,n}$. When not restricting the quantifier depth, we get $\Sigma_m = \bigcup_{n\geq 0} \Sigma_{m,n}$. Another way of thinking about~$\Sigma_m$ is to disallow negations and to allow both existential and universal quantifiers, starting with a possibly empty sequence of existential quantifiers; the index $m$ then corresponds to the number of blocks of quantifiers. Such formulae can be obtained by moving negations towards the atomic formulae, at which point they can easily be avoided. For $\Sigma_2$, we have the following property~\cite{diekertkufleitner2011tocs}:

\begin{lemma}\label{lem:alwaysopeninthealphabetictopology}
	If $L \subseteq A^\infty$ is definable in $\Sigma_2$, then $L$ is open in the alphabetic topology.
\end{lemma}

We also consider the fragment $\FO^2$ of first order formulae which uses only two different variables, say, $x$ and $y$. The fragment $\Sigma^2_{m,n}$ is the (syntactic) intersection of $\FO^2$ and $\Sigma_{m,n}$, i.e., $\varphi \in \Sigma^2_{m,n}$ if both $\varphi \in \FO^2$ and $\varphi \in \Sigma_{m,n}$. Similarly, $\Sigma^2_m$ is the intersection of $\FO^2$ and $\Sigma_m$. The emphasis on formulas (i.e., syntax) rather than the languages defined by the respective formulae (i.e., semantics) is important since, e.g., every $\FO^2$-definable language in $A^*$ is also definable in $\Sigma_2$; see~\cite{TherienWilke1998stoc}.

\paragraph{Relativization.} 

We write $\alpha \leq^{\Sigma^2}_{m,n} \beta$ for $\alpha,\beta \in A^\infty$ if the following implications holds for all sentences $\varphi \in \Sigma^2_{m,n}$:
\begin{equation*}
  \alpha \models \varphi \quad\Rightarrow\quad \beta \models \varphi
\end{equation*}
One of the key techniques used in this paper is relativization, in particular the results recorded in the following four lemmas.
The first lemma is about relativizing at the first $a \in A$; the second lemma is about relativizing between the last $b$ and the first $a$; and the last two are about relativizing at the last $b$.
 We only prove the last two lemmas; proofs for the first two over finite words can be found, e.g., in~\cite{FleischerKL17tocs}. The extension to infinite words is straightforward.

\begin{lemma}\label{lem:simplerelativization}
  Let $a \in A$. Suppose that $u a \alpha \leq^{\Sigma^2}_{m,n} v a \beta$ with $a \not\in \alp(u) \cup \alp(v)$. Then $u \leq^{\Sigma^2}_{m-1,n-1} v$ and $\alpha \leq^{\Sigma^2}_{m,n-1} \beta$.
\end{lemma}

\begin{lemma}\label{lem:squeezedrelativisation}
  Let $a,b \in A$. Suppose that $p b u a \alpha \leq^{\Sigma^2}_{m,n} q b v a \beta$ with $a \not\in \alp(pbu) \cup \alp(qbv)$ and $b \not\in \alp(ua\alpha) \cup \alp(va\beta)$. Then $u \leq^{\Sigma^2}_{m-1,n-1} v$.
\end{lemma}

\begin{lemma}\label{lem:dualrelativization}
  Let $b \in A$. Suppose that $u b \alpha \leq^{\Sigma^2}_{m,n} v b \beta$ with $b \not\in \alp(\alpha) \cup \alp(\beta)$. Then $u \leq^{\Sigma^2}_{m,n-1} v$ and $\alpha \leq^{\Sigma^2}_{m-1,n-1} \beta$.
\end{lemma}

\begin{proof}
  For a sentence $\varphi \in \Sigma^2_{m,n-1}$, we construct a sentence $\varphi_{<\mathsf{Y}b} \in \Sigma^2_{m,n}$ such that for all $w \in A^*$ and all $\gamma \in A^\infty$ with $b \not\in \alp(\gamma)$ we have $wb\gamma \vDash \varphi_{<\mathsf{Y}b}$ if and only if $w \vDash \varphi$.
  This then yields
  \begin{alignat*}{2}
    u \vDash \varphi 
    \quad \Rightarrow \quad &ub\alpha \vDash \varphi_{<\mathsf{Y}b} 
    &\qquad\quad&\text{by construction of $\varphi_{<\mathsf{Y}b}$}
    \\ \Rightarrow \quad &vb\beta \vDash \varphi_{<\mathsf{Y}b} 
    &&\text{since $u b \alpha \leq^{\Sigma^2}_{m,n} v b \beta$}
    \\ \Rightarrow \quad &v \vDash \varphi
    &&\text{by construction of $\varphi_{<\mathsf{Y}b}$}
  \end{alignat*}
  thereby showing $u \leq^{\Sigma^2}_{m,n-1} v$.
  We define $\varphi_{<\mathsf{Y}b}$ for arbitrary formulae $\varphi$ using structural induction; in particular, $\varphi$ can have free variables.  We maintain the following invariant:
	\begin{align*}
		&\text{For all $1 \leq p,q \leq \left|w\right|$, we have } wb\beta,p,q \vDash \varphi_{<\mathsf{Y}b} \text{ if and only if } w,p,q \vDash \varphi. 
	\end{align*}
	For the atomic formulae $\varphi$, we define $\varphi_{<\mathsf{Y}b} = \varphi$. The construction straightforwardly distributes over Boolean connectives; for instance, $(\varphi \wedge \psi)_{< \mathsf{Y}b} \,=\, \varphi_{< \mathsf{Y} b} \,\wedge\, \psi_{< \mathsf{Y} b}$. For existential quantification we define
	\begin{equation*}
		(\exists x \varphi)_{< \mathsf{Y} b} \,=\, \exists x \big( \left( \exists y\colon x < y  \wedge \lambda(y) = b \right) \wedge \varphi_{< \mathsf{Y}b} \big).
	\end{equation*}
	
	The approach to showing $\alpha \leq^{\Sigma^2}_{m-1,n-1} \beta$ is similar. For a formula $\varphi \in \Sigma^2_{m-1,n-1}$, we construct $\varphi_{> \mathsf{Y}b} \in \Sigma^2_{m,n}$ such that the following invariant holds for all $w \in A^*$ and all $\gamma \in A^\infty$ with $b \not\in \alp(\gamma)$:
	\begin{align*}
		&\text{For all $\left|wb\right| < p,q$, we have } wb\gamma,p,q \vDash \varphi_{>\mathsf{Y}b} \text{ if and only if } \gamma,p-\left|wb\right|,q-\left|wb\right| \vDash \varphi.
	\end{align*}
	Atomic formulae and Boolean connectives are as before. For existential quantification we define
	\begin{equation*}
		(\exists x \varphi)_{> \mathsf{Y} b} \,=\, \exists x \left( \left( \neg \exists y \colon x \leq y \wedge \lambda(y)=b \right) \wedge \varphi_{> \mathsf{Y}b} \right).
 		\qedhere
	\end{equation*}

\end{proof}

The following lemma discusses a similar situation as Lemma~\ref{lem:dualrelativization}, but with the first word having infinitely many occurrences of $b$ and the second word having only finitely many such occurrences. This is only relevant for $m \leq 2$ since for $m \geq 3$ one can define whether $b$ occurs infinitely often (resp.\ only finitely often) using the formula $\forall x \exists y\colon x<y \wedge \lambda(y) = b$ (resp.\ its negation).

\begin{lemma}\label{lem:dualrelativization2}
  Let $b \in A$. Suppose that $\alpha \leq^{\Sigma^2}_{m,n} v b \beta$ with $b \in \im(\alpha)$ and $b \not\in \alp(\beta)$. Then $\alpha \leq^{\Sigma^2}_{m,n-1} v$.
\end{lemma}

\begin{proof}
 	We use the construction of $\varphi_{<\mathsf{Y}b}$ from the proof of Lemma~\ref{lem:dualrelativization}.
	If $\alpha$ contains infinitely many $b$, then $\alpha \vDash \varphi$ if and only if $\alpha \vDash (\exists x \varphi)_{< \mathsf{Y} b}$. Thus, 
	\begin{equation*}
		\alpha \vDash \varphi \ \Rightarrow\ \alpha \vDash \varphi_{<\mathsf{Y}b} \ \Rightarrow\ vb\beta \vDash \varphi_{<\mathsf{Y}b} \ \Rightarrow\ v \vDash \varphi
	\end{equation*}
	thereby showing $\alpha \leq^{\Sigma^2}_{m,n-1} v$.
\end{proof}

\paragraph{Ehrenfeucht-Fra\"iss\'e games.}

We use Ehrenfeucht-Fra\"iss\'e (EF) games in order to determine if two words $\alpha$ and $\beta$ are indistinguishable in some fragment $\Sigma^2_{m,n}$. There are two players, Spoiler and Duplicator. By convention, Spoiler is male and Duplicator is female. The EF game $\Sigma^2_{m,n}(\alpha,\beta)$ consists of $n$ rounds. Spoiler and Duplicator are equipped with two pebbles each; typically, we assume that one pebble is labeled by $x$ and the other is labeled by $y$. In each round, Spoiler places a pebble on his words, thereby lifting an already placed pebble if necessary; Duplicator tries to copy the move on the other word using the pebble with the same label. After Duplicator's move, the two pebbles need to have the same labels and the same relative order on both words. Spoiler is allowed to change his word at most $m-1$ times. Initially, his word is $\alpha$, but he can use one of the $m-1$ word alternations before making his first move. Duplicator wins if she can copy (``duplicate'') each of the $n$ moves of Spoiler; otherwise Spoiler wins. Using standard techniques (see e.g.~\cite{str94}), it is straightforward to show that Duplicator has a winning strategy for $\Sigma^2_{m,n}(u,v)$ if and only if $u \leq^{\Sigma^2}_{m,n} v$.

\section{The Varieties $\Mvar{m}$}\label{sec:Mm}

The following section introduces the varieties $\Mvar{m}$ which we will use for characterizing the $\Sigma^{2}_{m}$-definable languages over infinite words. As one of our first results, we show that $\Mvar{m} = \intp{U_m \leq V_m} \cap \DA$; here, the right hand side is the family of varieties used for characterizing the languages definable in $\Sigma^{2}_{m}$ over finite words~\cite{FleischerKL17tocs}. The new way of defining these varieties is motivated by the proofs, in which we use a different way of climbing up the hierarchy of $\Sigma^{2}_{m}$-definable languages over infinite words.

\begin{definition}\label{def:KD}
  For a monoid $M$, let $u \preceq_{\mathbf{KD}} v$ for $u,v\in M$ if the following implications holds for all $s,t \in M$:
  \begin{enumerate}[(i)]
	\item\label{aaa:KD} If $s \Req svt$, then $s \Req sut$.
	\item\label{bbb:KD} If $svt \Leq t$, then $sut \Leq t$.
	\item\label{ccc:KD} If $s \Req sv$ and $vt \Leq t$, then $sut \leq svt$.
  \end{enumerate}
\end{definition}

\begin{lemma}\label{lem:KDstable}
	The relation $\preceq_{\mathbf{KD}}$ is a stable preorder.
\end{lemma}

\begin{proof}
	Reflexivity and transitivity are obvious. It remains to show that $\preceq_{\mathbf{KD}}$ is stable. Let $u \preceq_{\mathbf{KD}} v$. By left-right symmetry, it suffices to show $xu \preceq_{\mathbf{KD}} xv$. For \emph{(\ref{aaa:KD})}, suppose that $s \Req sxvt$. Then $s \Req sx$ and $sx \Req sxvt$. By $u \preceq_{\mathbf{KD}} v$, we obtain $sx \Req sxut$ and thus $s \Req sxut$. For \emph{(\ref{bbb:KD})}, suppose that $sxvt \Leq t$. By $u \preceq_{\mathbf{KD}} v$, we get $sxut \Leq t$ (we use $sx$ for $s$ in property \emph{(\ref{bbb:KD})}). Finally, for \emph{(\ref{ccc:KD})}, suppose that $s \Req sxv$ and $xvt \Leq t$. This yields $sx \Req sxv$ and $vt \Leq t$ and thus, by $u \preceq_{\mathbf{KD}} v$, we see that $sxut \leq sxvt$. This concludes the proof of $xu \preceq_{\mathbf{KD}} xv$.
\end{proof}

Let $u \equiv_{\mathbf{KD}} v$ if both $u \preceq_{\mathbf{KD}} v$ and $v \preceq_{\mathbf{KD}} u$. We point out the following immediate consequence of the definition of $\preceq_{\mathbf{KD}}$.

\begin{lemma}\label{lem:liftingEQ}
	If $s\Req sv$, $vt\Leq  t$ and $u \equiv_{\mathbf{K}\mathbf{D}} v$, then $sut = svt$.
\end{lemma}

A consequence of Lemma~\ref{lem:KDstable} is that $\equiv_{\mathbf{KD}}$ is a congruence. For a monoid $M$, we can therefore consider the quotient $\faktor{M}{\equiv_{\mathbf{KD}}}$ consisting of all congruence classes. Moreover, for $\equiv_{\mathbf{KD}}$-classes $[u]$ and $[v]$, we set $[u] \leq [v]$ if $u \preceq_{\mathbf{KD}} v$. This turns $\faktor{M}{\equiv_{\mathbf{KD}}}$ into an ordered monoid, denoted by $\faktor{M}{\preceq_{\mathbf{KD}}}$.
We note that, in general, $\faktor{M}{\preceq_{\mathbf{KD}}}$ is not an ordered quotient of $M$. That is, $u \leq v$ does not necessarily imply $[u] \leq [v]$ in $\faktor{M}{\preceq_{\mathbf{KD}}}$. For example, consider the multiplicative monoid $M = \left\{ 1,0 \right\}$ with the order $0 \leq 1$. We have that $1 \preceq_{\mathbf{KD}} 0$ since $0$ is the only element which is \greenJ-related to $0$. On the other hand, we do not have $0 \preceq_{\mathbf{KD}} 1$ because $1 \cdot 1 \Req 1$ while $1 \cdot 0 \not \Req 1$. This shows that $\faktor{M}{\preceq_{\mathbf{KD}}}$ is isomorphic to $\left\{ 1,0 \right\}$ with the order $1 \leq 0$. In other words, in this case, the quotient does nothing but changing the order.

\begin{lemma}\label{lem:SiinDA}
	For every monoid $M$, we have $M \in \DA$ if and only if $\faktor{M}{\preceq_{\mathbf{KD}}} \in \DA$.
\end{lemma}

\begin{proof}
	If $M \in \DA$, then $\faktor{M}{\preceq_{\mathbf{KD}}} \in \DA$ because $\faktor{M}{\preceq_{\mathbf{KD}}}$ is a (possibly unordered) divisor of $M$ and $\DA$ is closed under unordered division. For the other direction, suppose that $\faktor{M}{\preceq_{\mathbf{KD}}} \in \DA$. Consider $x,y,z \in M$. We have 
	\begin{equation*}
	  (xyz)^{\omega}y(xyz)^{\omega} 
	  \equiv_{\mathbf{KD}} 
	  (xyz)^{\omega}
	\end{equation*}
	Applying Lemma~\ref{lem:liftingEQ} with $s=t=(xyz)^\omega$ yields
	\begin{align*}
	  (xyz)^{\omega}y(xyz)^{\omega}
	  &=
	  (xyz)^{\omega} \cdot (xyz)^{\omega}y(xyz)^{\omega} \cdot (xyz)^{\omega}
	  \\ &= 
	  (xyz)^{\omega} \cdot (xyz)^{\omega} \cdot (xyz)^{\omega}
	  =
	  (xyz)^{\omega}
	\end{align*}
	This shows $M \in \DA$.
\end{proof}

\begin{definition}
	Let $\Mvar{1} = \intp{1 \leq z}$. For $m \geq 2$, let $\Mvar{m}$ be the class of all finite ordered monoids~$M$ such that $\faktor{M}{\preceq_{\mathbf{KD}}} \in \Mvar{m-1}$.
\end{definition}

Since $\mathbf{J}^+ \subseteq \DA$, Lemma~\ref{lem:SiinDA} shows that $\Mvar{m} \subseteq \DA$ for all $m \geq 1$. If $M \in \DA$, then $1 \leq v$ implies $1 \preceq_{\mathbf{KD}} v$ (properties \emph{(\ref{aaa:KD})} and \emph{(\ref{bbb:KD})} in Definition~\ref{def:KD} hold by Lemma~\ref{lem:alphabetLR}; property \emph{(\ref{ccc:KD})} is trivial). This shows that if $M \in \intp{1 \leq z}$, then $\faktor{M}{\preceq_{\mathbf{KD}}} \in \intp{1 \leq z}$. Thus, $\Mvar{m} \subseteq \Mvar{m+1}$ for all $m \geq 1$.

Next, we show that each class $\Mvar{m}$ can be defined using $\omega$-terms. To this end, we introduce a family of $\omega$-terms $U_m$, $V_m$ for $m \geq 1$. Each of the $\omega$-terms $U_m$ and $V_m$ for $m \geq 2$ uses variables $x_2,\ldots,x_m,y_2,\ldots,y_m,z$. For $m = 1$, let $U_1 = 1$ and $V_1 = z$. For $m \geq 2$, let
\begin{align*}
	U_m & \,=\, \left( V_{m-1} x_m \right)^{\omega}U_{m-1}\left( y_m V_{m-1} \right)^{\omega} \\
	V_m & \,=\, \left( V_{m-1} x_m \right)^{\omega}V_{m-1}\left( y_m V_{m-1} \right)^{\omega}
\end{align*}

\begin{proposition}\label{prp:intersectionsoftwh}
	We have \,$\Mvar{m} = \intp{U_m \leq V_m} \cap \DA$\, for all $m \geq 1$.
\end{proposition}

\begin{proof}
	We proceed by induction. The statement for $m=1$ holds by definition. Let now $m \geq 2$. We first assume that $M \in \Mvar{m}$. By Lemma \ref{lem:SiinDA}, we have $M \in \DA$. Furthermore, let $x_2,\ldots,x_m,y_2,\ldots,y_m,z \in M$. These elements define $U_{m-1}$ and $V_{m-1}$ satisfying  $U_{m-1} \preceq_{\mathbf{KD}} V_{m-1}$ by induction. We have $(V_{m-1}x_{m})^{\omega}V_{m-1} \cdot x_m (V_{m-1}x_{m})^{\omega-1} = (V_{m-1}x_{m})^{\omega}$
	and hence $(V_{m-1}x_{m})^{\omega}V_{m-1} \Req \left( V_{m-1} x_{m} \right)^{\omega}$. Similarly, we see that $V_{m-1} \left( y_{m} V_{m-1}  \right)^{\omega} \Leq  (y_{m}V_{m-1})^{\omega}$. It follows that $U_{m} \leq V_{m}$ by definition of $\preceq_{\mathbf{KD}}$. Since this holds for arbitrary elements $x_2,\ldots,x_m,y_2,\ldots,y_m,z \in M$, we have $M \in \intp{U_m \leq V_m}$.

	For the other direction, let $M \in \intp{U_m \leq V_m} \cap \DA$. Let $x_2,\ldots,x_{m-1},y_2,\ldots,y_{m-1},z \in M$ be arbitrary and consider $U_{m-1}, V_{m-1} \in M$ defined by these elements. We claim that $U_{m-1} \preceq_{\mathbf{KD}} V_{m-1}$. Since $M \in \DA$, Lemma~\ref{lem:alphabetLR} shows that the following two implications hold for all $s,t \in M$:
	\begin{enumerate}[\itshape(i)]
	\item If $s \Req sV_{m-1}t$, then $s \Req sU_{m-1}t$.
	\item If $sV_{m-1}t \Leq t$, then $sU_{m-1}t \Leq t$.
    \end{enumerate}
	For property \emph{(\ref{ccc:KD})} in the definition of $\preceq_{\mathbf{KD}}$, suppose that $sV_{m-1} \Req s$ and $V_{m-1}t \Leq t$. Choose $x_m,y_m \in M$ such that  $sV_{m-1}x_m = s$ and $y_mV_{m-1}t$. Since $M \in \intp{U_m \leq V_m}$, we get 
	\begin{align*}
		sU_{m-1}t 
		&= s (V_{m-1} x_m)^{\omega} U_{m-1} (y_m V_{m-1})^{\omega} t
		\\ &\leq s(V_{m-1} x_m)^{\omega} V_{m-1} (y_m V_{m-1} )^{\omega}t
		= sV_{m-1}t.
    \end{align*}
    This concludes the proof of the claim $U_{m-1} \preceq_{\mathbf{KD}} V_{m-1}$. Therefore, we have $\faktor{M}{\preceq_{\mathbf{KD}}} \in \intp{U_{m-1} \leq V_{m-1}} \cap \DA$. Induction yields $\faktor{M}{\preceq_{\mathbf{KD}}} \in \Mvar{m-1}$. Hence, by definition of $\Mvar{m}$, we see that $M \in \Mvar{m}$.
\end{proof}

This in particular shows that each $\Mvar{m}$ forms a positive variety.
We now state the main result of this paper.

\begin{theorem}\label{thm:sigmatwo}
  Let $L \subseteq A^{\infty}$ be an $\omega$-regular language and let $\mu: A^* \to M$ be its syntactic homomorphism. The following properties hold:
  \begin{enumerate}[(i)]
  \item\label{aaa:sigmatwo} $L$ is definable in $\Sigma^2_{1}$ if and only if both $M \in \Mvar{1}$ and $L$ is open in the Cantor topology.
  \item\label{bbb:sigmatwo} $L$ is definable in $\Sigma^2_{2}$ if and only if both $M \in \Mvar{2}$ and $L$ is open in the alphabetic topology.
  \item\label{ccc:sigmatwo} If $m \geq 3$, then $L$ is definable in $\Sigma^2_{m}$ if and only if $M \in \Mvar{m}$.
  \end{enumerate}
\end{theorem}

The proof of Theorem~\ref{thm:sigmatwo} is given in the next section.
In order to give a version of Theorem~\ref{thm:sigmatwo} for $A^\omega$, we need to introduce the respective counter-parts of some definitions.
For a formula $\varphi$ in first-order logic, let $L_{\omega}(\varphi) = \left\{ \alpha \in A^\omega \mid \alpha \vDash \varphi \right\}$. We have $L_{\omega}(\varphi) = L(\varphi) \cap A^\omega$. We say that a language $L \subseteq A^\omega$ is \emph{$\omega$-definable} in a fragment $\mathcal{F}$ if there exists $\varphi \in \mathcal{F}$ such that $L = L_{\omega}(\varphi)$.
The Cantor topology and the alphabetic topology have natural counterparts in $A^\omega$: a basis for the \emph{$\omega$-Cantor topology} is $\left\{ uA^{\omega} \mid u \in A^* \right\}$ and a basis for the \emph{$\omega$-alphabetic topology} is $\left\{ uB^{\omega} \mid u \in A^* \text{ and } \emptyset \neq B \subseteq A \right\}$. This allows us to formulate the following version of Theorem~\ref{thm:sigmatwo} for infinite words $A^{\omega}$ (rather than $A^{\infty}$).

\begin{corollary}\label{cor:sigmatwoomega}
  Let $L \subseteq A^{\omega}$ be an $\omega$-regular language and let $\mu: A^* \to M$ be its syntactic homomorphism. The following properties hold:
  \begin{enumerate}[(i)]
  \item\label{aaa:sigmatwoomega} $L$ is $\omega$-definable in $\Sigma^2_{1}$ if and only if both $M \in \Mvar{1}$ and $L$ is open in the $\omega$-Cantor topology.
  \item\label{bbb:sigmatwoomega} $L$ is $\omega$-definable in $\Sigma^2_{2}$ if and only if both $M \in \Mvar{2}$ and $L$ is open in the $\omega$-alphabetic topology.
  \item\label{ccc:sigmatwoomega} If $m \geq 3$, then $L$ is $\omega$-definable in $\Sigma^2_{m}$ if and only if $M \in \Mvar{m}$.
  \end{enumerate}
\end{corollary}

\begin{proof}
  For \emph{(\ref{ccc:sigmatwoomega})}, there remains nothing to show because the language $A^\omega$ (as a subset of $A^\infty$) is definable in $\Sigma^2_3$. We now prove \emph{(\ref{aaa:sigmatwoomega})} and \emph{(\ref{bbb:sigmatwoomega})}.
  We first consider the implications from left to right. Suppose that $L = L_{\omega}(\varphi)$ for $\varphi \in \Sigma^2_m$. Consider $L' = L(\varphi) \subseteq A^\infty$. By Theorem~\ref{thm:sigmatwo}, there exists a homomorphism $\mu' : A^* \to M'$ with $M' \in \Mvar{m}$ which recognizes $L'$. Moreover, we can assume that $\mu'$ is alphabetic (see Example~\ref{exa:alp}).
  Using the notation $[s] = {\mu'\hspace*{1pt}}^{-1}(s)$, we have 
  \begin{align*}
    L 
    &= L' \cap A^\omega 
    \\ &= \bigcup \left\{ [s][e]^{\omega} \mid (s,e) \text{ is a linked pair and } [s][e]^{\omega} \cap L' \neq \emptyset \text{ and } e \neq 1\right\}
    \\ &= \bigcup \left\{ [s][e]^{\omega} \mid (s,e) \text{ is a linked pair and } [s][e]^{\omega} \cap L \neq \emptyset \right\}
  \end{align*}
  Moreover, $L$ is upward closed with respect to $\mu'$. Therefore, $M'$ recognizes $L$. Since $M$ is a divisor of $M'$ and since $\Mvar{m}$ is a variety, we have $M \in \Mvar{m}$. If $m=1$, then $L'$ is open in the Cantor topology; hence, $L$ is open in the $\omega$-Cantor topology. Similarly, if $m=2$, then $L'$ is open in the alphabetic topology; thus, $L$ is open in the $\omega$-alphabetic topology.

  Next, consider the implications from right to left. Let 
  \begin{align*}
    L' = L \,\cup\, \bigcup \left\{ [s] \mid (s,e) \text{ is a linked pair and } [s][e]^{\omega} \cap L \neq \emptyset\right\}
  \end{align*}
  By construction, $L = L' \cap A^\omega$ and $L' \subseteq A^\infty$ is recognized by $M$. Moreover, if $L$ is open in the $\omega$-Cantor topology, then $L'$ is open in the Cantor topology. Similarly, if $L$ is open in the $\omega$-alphabetic topology, then $L$ is open in the alphabetic topology. Thus, by Theorem~\ref{thm:sigmatwo}, there exists $\varphi \in \Sigma^2_m$ with $L' = L(\varphi)$. It follows that $L_\omega(\varphi) = L(\varphi) \cap A^\omega = L' \cap A^\omega = L$. This shows that $L$ is $\omega$-definable in $\Sigma^2_m$.
\end{proof}

Similar to Corollary~\ref{cor:sigmatwoomega}, in order to show that Theorem~\ref{thm:sigmatwo} implies the characterization of $\Sigma^2_m$ over finite words, we first introduce the notion of definability for finite words. For a formula $\varphi$ in first-order logic, let $L_{*}(\varphi) = \left\{ \alpha \in A^* \mid \alpha \vDash \varphi \right\}$. Note that $L_{*}(\varphi) = L(\varphi) \cap A^*$. We say that a language $L \subseteq A^*$ is \emph{$*$-definable} in a fragment $\mathcal{F}$ if there exists $\varphi \in \mathcal{F}$ such that $L = L_{*}(\varphi)$.

\begin{corollary}[Fleischer, Kuf\-leitner, Lauser \cite{FleischerKL17tocs}]\label{cor:sigmatwostar}
  A language $L \subseteq A^*$ is $*$-definable in $\Sigma^2_{m}$ if and only if its syntactic homomorphism $\mu: A^* \to M$ satisfies $M \in \Mvar{m}$.
\end{corollary}

\begin{proof}
  There is nothing to show for $m \geq 2$ because the language $A^*$ (as a subset of $A^\infty$) is definable in $\Sigma^2_2$. Therefore, we only consider the case $m=1$. Here, the characterization is well-known; see e.g.~\cite{diekertgastinkufleitner2008ijfcs}. For completeness, we show how to derive it from the result for $A^\infty$ using standard techniques.
  
  For the implication from left to right, suppose that $L = L_{*}(\varphi)$ for $\varphi \in \Sigma^2_1$. Consider $L' = L(\varphi) \subseteq A^\infty$. By Theorem~\ref{thm:sigmatwo}, there exists a homomorphism $\mu' : A^* \to M'$ with $M' \in \Mvar{1} = \intp{1 \leq z}$ which recognizes $L'$. Moreover, we can assume that $\mu'$ is alphabetic (see Example~\ref{exa:alp}).
  Using the notation $[s] = {\mu'\hspace*{1pt}}^{-1}(s)$, we have $[s] = [s][(1,\emptyset)]^\omega$ and thus
  \begin{align*}
    L 
    \;=\; L' \cap A^*
    \;=\; \bigcup_{[s] \cap L' \neq \emptyset} \hspace*{-5pt} [s]
    \;=\; \bigcup_{[s] \cap L \neq \emptyset} \hspace*{-4pt} [s]
  \end{align*}
  This shows that $L$ is recognized by $M' \in \intp{1 \leq z}$. Since $M$ divides $M'$, it follows that $M \in \intp{1 \leq z}$. 
  
  For the implication from right to left, suppose that $M \in \intp{1 \leq z}$. Let $L' = L A^\infty$. We have $\mu(u) = \mu(u) \cdot 1 \leq \mu(u)\, \mu(v) = \mu(uv)$ for all $u,v \in A^*$. Therefore, if $u \in L$, then by upward closure we have $uv \in L$ for all $v \in A^*$. This shows that $L = LA^*$. Therefore, $L = L' \cap A^*$ and $L'= \bigcup \left\{ [s][e]^{\omega} \mid (s,e) \text{ is a linked pair and } [s] \cap L \neq \emptyset \right\}$; in particular, $L'$ is open in the Cantor topology and recognized by $M$. By Theorem~\ref{thm:sigmatwo}, there exists $\varphi \in \Sigma^2_1$ such that $L' = L(\varphi)$. By construction of $L'$, we have $L = L_{*}(\varphi)$.
\end{proof}

Since the syntactic homomorphism is effectively computable from all major presentations of $\omega$-regular languages (e.g., automata, $\omega$-regular expressions or MSO logic), Theorem~\ref{thm:sigmatwo} together with the decidability of the two topologies gives the following corollary. Similar results hold for $A^*$ and $A^\omega$.

\begin{corollary}
	For all $m \geq 1$, it is decidable whether a given language $L \subseteq A^{\infty}$ is definable in $\Sigma^2_m$.
\end{corollary}

\section{Decidability of $\Sigma^2_m$}

This section is devoted to the proof of Theorem~\ref{thm:sigmatwo}. Every $\FO^2$ definable language over finite words is definable in $\Sigma_2$; see~\cite{TherienWilke1998stoc}. Over infinite words, this inclusion still holds for the languages which are open in the alphabetic topology~\cite{diekertkufleitner2011tocs}. In particular, this inclusion (relative to the open sets of the alphabetic topology) holds for all fragments $\Sigma^2_m$. For $m=1$ however, the fragments $\Sigma^2_1$ and $\Sigma_1$ have the same expressive power. The corresponding characterizations over $A^*$ and $A^\omega$ are well known; see e.g.~\cite{diekertgastinkufleitner2008ijfcs,PerrinPin2004}. For completeness we give a self-contained proof for the characterization of $\Sigma^2_1$ and, hence, of $\Sigma_1$ over $A^\infty$.

\begin{proof}[Proof of Theorem~\ref{thm:sigmatwo}\,{(\ref{aaa:sigmatwo})}]
	First, assume that the syntactic monoid of $L$ is in $\intp{1 \leq z}$ and that $L$ is open in the Cantor topology. Let $U$ be the set of finite words in $L$. Since $L$ is open, every word $\alpha \in L$ has a finite prefix $u \in L$. Thus, $L = UA^\infty$. According to Higman's Lemma (see e.g.\ \cite[Theorem 6.13]{DiekertEtAl2016}), there exists a finite set of words $u_1, \dots , u_k \in U$ which are minimal with respect to the subword relation, i.e., every word $u \in U$ has some $u_i$ as a subword. Therefore, every word in $L$ has some $u_i$ as a subword. Conversely, if $\alpha$ has some $u_i$ as a subword, then $\alpha \in L$ because the syntactic monoid satisfies $1 \leq z$. For $u = a_1 \cdots a_{\ell}$, let $\varphi_u$ be the formula
	\begin{equation*}
		\exists x ( \lambda(x) = a_1 \wedge \exists y ( x < y \wedge \lambda(y) = a_2 \wedge \exists x ( y < x \wedge \lambda(x) = a_3 \wedge \exists y( \cdots ))))
	\end{equation*}
	We have $\alpha \vDash \varphi_u$ if and only if $u$ is a subword of $\alpha$. It follows that $L$ is defined by the $\Sigma^2_1$-formula $\bigvee_i \varphi_{u_i}$.

	For the other direction, we first note that the language is $\omega$-regular since it is, in particular, definable in $\FO^2$. Furthermore, for every $\Sigma^2_1$-formula $\varphi$ we have that $xyw^{\omega} \vDash \varphi$ implies $xzyw^{\omega} \vDash \varphi$, and $xy^{\omega} \vDash \varphi$ implies $x(zy)^{\omega} \vDash \varphi$. So we have $1 \leq z$ in the syntactic monoid. Finally, every word $\alpha$ satisfying $\varphi \in \Sigma^2_1$ has a finite prefix $u$ with $u \vDash \varphi$ and hence $\alpha \in u A^\infty \subseteq L(\varphi)$. Therefore, $L(\varphi)$ is open in the Cantor topology.
\end{proof}

\subsection{From Logic to Algebra}

The proofs of both \emph{(\ref{bbb:sigmatwo})} and \emph{(\ref{ccc:sigmatwo})} in Theorem~\ref{thm:sigmatwo} use similar techniques which are introduced below. The direction from logic to algebra relies on EF games. The following lemma introduces a standard technique for EF games; it shows that winning strategies behave well with respect to both finite and infinite concatenation.

\begin{lemma}\label{lem:omegacongruence}
	Let $A$ be an alphabet and let $u, v \in A^*$ and $\alpha,\beta \in A^{\infty}$. Suppose that Duplicator has winning strategies for $\Sigma^2_{m,n}(u,v)$ and $\Sigma^2_{m,n}(\alpha,\beta)$. Then Duplicator has a winning strategy for $\Sigma^2_{m,n}(u\alpha,v\beta)$. Similarly, if $u_1,u_2,u_3,\ldots$ and $v_1,v_2,v_3,\ldots$ are sequences of finite words such that Duplicator has a winning strategy for $\Sigma^2_{m,n}(u_i,v_i)$ for all $i$, then Duplicator also has a winning strategy for $\Sigma^2_{m,n}\left( u_1 u_2 u_3 \cdots, v_1 v_2 v_3 \cdots \right)$.
\end{lemma}

\begin{proof}
  If Duplicator has a winning strategy for $\Sigma^2_{m,n}(\alpha,\beta)$, then this winning strategy also works for all games $\Sigma^2_{m',n'}(\alpha,\beta)$ with $m' \leq m$ and $n' \leq n$; similarly, Duplicator also wins $\Sigma^2_{m',n'}(\beta,\alpha)$ for all $m' < m$ and $n' \leq n$ by applying this strategy. 
  
  Therefore, if Duplicator wins both $\Sigma^2_{m,n}(u,v)$ and $\Sigma^2_{m,n}(\alpha,\beta)$, then she also wins the game $\Sigma^2_{m,n}(u\alpha,v\beta)$ by applying the following strategy: Whenever Spoiler makes a move on either $u$ or $v$, then Duplicator responds according to her strategy for $\Sigma^2_{m,n}(u,v)$; and if Spoiler makes a move on either $\alpha$ or $\beta$, then Duplicator responds according to her strategy for $\Sigma^2_{m,n}(\alpha,\beta)$. Note that Spoiler has no advantage by changing between the words on the left $u,v$ and the words on the right $\alpha,\beta$ because all left positions are smaller than all right positions. Therefore, the relative orders of the $x$- and  the $y$-pebbles are identical if they are placed on different sides (left or right) such that the sides are the same for both players.
  
  The second part of the lemma is similar: Duplicator wins by applying the winning strategy for the game $\Sigma^2_{m,n}(u_i,v_i)$ whenever Spoiler made his last move within $u_i$ or $v_i$.
\end{proof}

Suppose that Duplicator has a winning strategy for $\Sigma^2_{m,n}(u,v)$. The last part of Lemma~\ref{lem:omegacongruence} implies that she has a winning strategy for $\Sigma^2_{m,n}(u^{\omega},v^{\omega})$.

\begin{lemma}\label{lem:efgame}
	Let $m \geq 2$, $n \geq 1$ and let $u, v \in A^*$. If Duplicator has a winning strategy for $\Sigma^2_{m-1,n}(u,v)$, then she has a winning strategy for $\Sigma^2_{m,n}(p^nuq^n, p^nvq^n)$ for all $p,q \in A^*$ such that $\alp(v) \subseteq \alp(p) \cap \alp(q)$.
\end{lemma}

\begin{proof}
  Both words $p^nuq^n$ and $p^nvq^n$ have a \emph{$p$-block} $p^n$ on the left and a \emph{$q$-block} $q^n$ on the right; the \emph{center} is either $u$ or $v$. By Lemma~\ref{lem:omegacongruence}, there exists a winning strategy for Duplicator for $\Sigma^2_{m-1,n}(p^nuq^n, p^nvq^n)$ such that whenever Spoiler places a pebble in one of the blocks, then Duplicator copies the move in the same block of the other word; if Spoiler places his pebble in the center, then Duplicator responds according to her strategy for $\Sigma^2_{m-1,n}(u,v)$. 
  
  We now describe Duplicators winning strategy for the remaining game. We can assume that Spoiler never moves the same pebble twice in a row (otherwise, Duplicator could simply respond as if Spoiler had not made the first of the two moves). Without loss of generality, we assume that Spoiler makes his remaining moves on $p^nvq^n$; the proof for the other word is similar since $u \leq^{\Sigma^2}_{m-1,n} v$ yields $\alp(u) \subseteq \alp(v)$ and thus $\alp(u) \subseteq \alp(p) \cap \alp(q)$. The remaining game is played in rounds, starting with round $1$. By abuse of notation, we say that round $0$ is the situation after Duplicator played her winning strategy for the game $\Sigma^2_{m-1,n}(p^nuq^n, p^nvq^n)$.
  We show that Duplicator has a winning strategy for the remaining game which, after every round, satisfies the following invariant: If, in round $k$, Spoiler places his pebble on position $i$ and Duplicator places her pebble on position $j$, then we require that:
	\begin{enumerate}[\itshape(a)]
	\item\label{aaa:efgamecondition} If $i \leq |p^{n-k}|$, then $j = i$.
	\item\label{bbb:efgamecondition} If $|p^{n-k}| < i \leq |p^n|$, then $|p^{n-k}| < j \leq i$.
	\item\label{ccc:efgamecondition} If $|p^n| < i \leq |p^n v|$, then $|p^{n-k}| < j \leq |p^n u q^k|$.
	\item\label{ddd:efgamecondition} If $|p^n v| < i \leq |p^n v q^k|$, then  $i - |v| + |u| \leq j \leq |p^n v q^k|$.
	\item\label{eee:efgamecondition} If $i > |p^nvq^k|$, then $i - |v| + |u| = j$.
	\end{enumerate}
	In other words, outside the central factors $p^k u q^k$ and $p^k v q^k$, Duplicator always copies Spoiler's move; and within these factors, Duplicator never moves her pebble closer to the center $u$ than Spoiler to $v$. Note that the invariant is satisfied after round $0$.
	
	Consider the round $k>0$ and suppose that the pebbles in the previous round were placed on position~$i'$ in $p^n v q^n$ and position $j'$ in $p^n u q^n$ such that the invariant for round $k-1$ is satisfied (or that no pebbles have yet been placed). We can assume that Spoiler places his pebble on position $i$ with $i < i'$. (The case $i > i'$ is symmetric and the case $i=i'$ is trivially answered by $j=j'$; if the positions $i'$ and $j'$ do not exist because no pebbles have yet been placed, then we can ignore the verification of $j < j'$ in the description of the strategy below.)
	We distinguish three cases:
	\begin{enumerate}[\itshape(i)]
	\item\label{aaa:efgame} $i \leq |p^{n-k}|$
	\item\label{bbb:efgame} $|p^{n-k}| < i \leq |p^n v|$
	\item\label{ccc:efgame} $|p^n v| < i$
	\end{enumerate}
	In case \textit{(\ref{aaa:efgame})}, Duplicator moves her pebble to the position $j = i$ on the word $p^n u q^n$. In both cases $i' \leq |p^{n-k+1}|$ and $i' > |p^{n-k+1}|$, we have $j < j'$ (either because $j = i < i' = j'$ or because $j < |p^{n-k+1}| < j'$, respectively). Therefore, this is a legal move for Duplicator. Moreover, this choice of $j$ trivially satisfies the invariant.
	
	In case \textit{(\ref{bbb:efgame})}, Duplicator moves her pebble to the leftmost position $j > |p^{n-k}|$ in $p^n u q^n$ which has the same label as $i$. Such a position exists since $\alp(v) \subseteq \alp(p)$. Note that $j \leq |p^{n-k+1}|$ and $j \leq i$. In particular, the invariant for round $k$ is satified. To show that this choice of $j$ yields a legal move for Duplicator, we again need to distinguish the two cases $i' \leq |p^{n-k+1}|$ and $i' > |p^{n-k+1}|$. In both cases, we see that $j < j'$ (either because $j \leq i < i' = j'$ or because $j \leq |p^{n-k+1}| < j'$, respectively).
	
	Finally, we consider case \textit{(\ref{ccc:efgame})}. Duplicator chooses $j = i - |v| + |u|$. Then $j < i' - |v| + |u| \leq j'$. Thus, this yields a legal move for Duplicator which satisfies the invariant.
	
	
	This shows that Duplicator has a winning strategy for at least $n$ rounds in the remaining game.
\end{proof}

By Lemma~\ref{lem:omegacongruence}, the relation $\leq^{\Sigma^2}_{m,n}$ is a stable partial order on $A^*$ for all $m$, $n$. We consider the monoid $N_{m,n} = \faktor{A^*}{\leq^{\Sigma^2}_{m,n}}$. It is a standard result from finite model theory that there are only finitely many inequivalent formulae in $\Sigma^2_{m,n}$; see e.g.~\cite[Prop.~IV.1.1]{str94}. Thus, $N_{m,n}$ is finite. The next lemma shows that $N_{m,n}$ recognizes all languages definable in $\Sigma^2_{m,n}$. Note that this is non-trivial because we consider languages in $A^{\infty}$ while $N_{m,n} = \faktor{A^*}{\leq^{\Sigma^2}_{m,n}}$ only considers equivalence over finite words.

\begin{lemma}\label{lem:nmnRecL}
	If $L$ is definable in $\Sigma^2_{m,n}$, then $N_{m,n}$ recognizes $L$.
\end{lemma}

\begin{proof}
	Suppose that $L = L(\varphi)$ for $\varphi \in \Sigma^{2}_{m,n}$, and let $\nu: A^* \to N_{m,n}$ be the natural projection. Let
	\begin{equation*}
	  P \,=\, \bigcup \left\{ [s][e]^{\omega} \mid (s,e) \text{ is a linked pair and } [s][e]^{\omega} \cap L \neq \emptyset \right\}
	\end{equation*}
	By Lemma~\ref{lem:omegacongruence}, we see that $P \subseteq L$ and that $L$ is upward closed. It remains to show that $L \subseteq P$. Consider a word $\alpha \in L$. There exists a linked pair $(s,e)$ over $N_{m,n}$ such that $\alpha \in [s][e]^\omega$. In particular, we have $\alpha \in [s][e]^\omega \cap L \neq \emptyset$. By definition of $P$, we have $[s][e]^\omega \subseteq P$ and, hence, $\alpha \in P$.
\end{proof}

\begin{lemma}\label{lem:nmnIdentity}
	For every $n,m \in \mathbb{N}$, we have $N_{m,n} \in \intp{U_m \leq V_m}$.
\end{lemma}

\begin{proof}
	We proceed by induction on $m$.
	Fix $x_2, \dots, x_m, y_2, \dots, y_m, z \in A^*$. For an integer  $k \geq 0$, let $U_{1,k} = \varepsilon$, $V_{1,k} = z$ and
	\begin{align*}
		U_{i,k} & = (V_{i-1,k}\,x_i)^{k} \, U_{i-1,k} \, (y_iV_{i-1,k})^{k}
		\\ V_{i,k} & = (V_{i-1,k}\,x_i)^{k} \, V_{i-1,k} \, (y_iV_{i-1,k})^{k}
	\end{align*}
	In other words, $U_{i,k}$ and $V_{i,k}$ are the $\omega$-terms $U_i$ and $V_i$, respectively, with words substituted for the variables and the formal $\omega$-powers replaced by the integer $k$.
	We claim that $U_{m,k} \leq^{\Sigma^2}_{m,n} V_{m,k}$ for all $k \geq n$. This holds for $m = 1$ because Spoiler cannot make any moves on $U_{1,k} = \varepsilon$. Next, assume that the claim holds for $m-1$. Then Duplicator has a winning strategy for $\Sigma^2_{m-1,n}(U_{m-1,k},V_{m-1,k})$. By Lemma~\ref{lem:efgame}, she has a winning strategy for $\Sigma^2_{m,n}(U_{m,k},V_{m,k})$. Thus, $U_{m,k} \leq^{\Sigma^2}_{m,n} V_{m,k}$ as desired.
	
	Let $\nu: A^* \to N_{m,n}$ be the natural projection and let $k \geq n$ be a multiple of $\omega_{N_{m,n}}$. Suppose that $u = I(U_m)$ and $v = I(V_m)$ for some interpretation $I$ of $\omega$-terms. We can choose words $x_2,\dots,x_m,y_2,\dots,y_m,z \in A^*$ such that $u = \nu(U_{m,k})$ and $v = \nu(V_{m,k})$. Since $U_{m,k} \leq^{\Sigma^2}_{m,n} V_{m,k}$, we have $u \leq v$ in $N_{m,n}$. Hence, $N_{m,n} \in \intp{U_m \leq V_m}$.
\end{proof}

We can now combine the previous lemmas to obtain the direction from logic to algebra in Theorem~\ref{thm:sigmatwo}.

\begin{proposition}\label{prp:logic2algebra}
  If $L \subseteq A^\infty$ is definable in $\Sigma^2_m$, then the syntactic monoid of $L$ is in $\Mvar{m}$.
\end{proposition}

\begin{proof}
  Let $M$ be the syntactic monoid of $L$.
  By Lemma~\ref{lem:nmnRecL}, there exists $n$ such that $L$ is recognized by $N_{m,n}$. By Lemma~\ref{lem:nmnIdentity}, we have $N_{m,n} \in \intp{U_m \leq V_m}$. Hence, $M \in \intp{U_m \leq V_m}$. Since $L$ is definable in $\FO^2$, we have $M \in \DA$; see~\cite{diekertkufleitner2011tocs}. By Proposition~\ref{prp:intersectionsoftwh}, we conclude $M \in \Mvar{m}$.
\end{proof}

\subsection{From Algebra to Logic}

We now turn to the other direction, where we will use two lemmas, Lemma \ref{lem:mainfactorizationlemmaIM} and \ref{lem:mainfactorizationlemmaN}, which are adaptations of a lemma used in the finite case. We consider two words $\alpha \leq^{\Sigma^2}_{m,n} \beta$ for a particular $n$, and give useful factorizations for these words. The novelty in our adaptation is the treatment of the infinite part. A key element for this is the concept of a last finitely occurring letter. Whenever $m \geq 3$, formulae of the form
\begin{equation*}
  \exists x (\lambda(x) = a \wedge \neg \exists y : y > x \wedge  \lambda(y) \in B)
\end{equation*}
and negations thereof show that the last finitely occurring letter and the imaginary alphabet are the same in $\alpha,\beta$ as soon as $\alpha \leq^{\Sigma^2}_{3,2} \beta$. Here, $\lambda(y) \in B$ is a macro for $\bigvee_{b\in B} \lambda(y)=b$. However, if $m = 2$, we cannot rely on this property.

The proof of Lemmas \ref{lem:mainfactorizationlemmaIM} and \ref{lem:mainfactorizationlemmaN} will use so-called \greenR- and \greenL-factorizations, which we will now define. These are standard concepts for finite words; see e.g.~\cite{kufleitnerweil2012csl}. Here we introduce some generalizations. First we define a suitable $\greenL$-factorization for $A^\infty$. This is done by starting the factorization at the last finitely occurring letter.

Consider a homomorphism $\mu: A^* \to M$. Let $\alpha = u\alpha'$ where $u$ is the shortest prefix such that $\im(\alpha) = \alp(\alpha')$. There exists a linked pair $(s,e)$ over~$M$ such that $\alpha' \in [s][e]^{\omega}$. The \emph{\greenL-factorization of $\alpha$} is the unique factorization $u_1 b_1 \dots u_n b_n \alpha'$ with $u=u_1 b_1 \dots u_n b_n$ where $b_n$ is the last finitely occurring letter of $\alpha$ (we have $u=\varepsilon$ if such a letter does not exist), $\mu(u_n b_n) s \Leq \mu(b_n) s$ and 
	\begin{equation*}
		\mu(u_i b_i u_{i+1} \dots u_n b_n) s \Leq \mu(b_i u_{i+1} \dots u_n b_n) s \Ll \mu(u_{i+1} \dots u_n b_n) s
	\end{equation*}
for all $1 \leq i \leq n-1$.
In order to see that this factorization is well defined, we want to show that it does not depend on the choice of $(s,e)$. For this, assume that $\alpha \in [t][f]^\omega$ for a linked pair $(t,f)$. This implies $(s,e) \sim (t,f)$; see e.g.~\cite[Corollary~II.2.9]{PerrinPin2004}. Thus, there exists~$x$ such that $t = sx$. If $ws \Leq s$, there exists $v$ such that $vws = s$. Multiplying to the right by $x$ shows $vwt = t$ and, thus, $wt \Leq t$. In particular, for all $w$ we have $ws \Leq s$ if and only if $wt \Leq t$. This shows that the relations above do not depend on $s$, making the \greenL-factorization well defined. Note that for finite words (with the possible exception of $b_n$) this definition coincides with the usual $\greenL$-factorization.

Consider a homomorphism $\mu: A^* \to M \in \DA$. We construct the \greenR-factorization for a pair of (finite or infinite) words $(\alpha,\beta)$ with respect to $\beta$. Over finite words, this factorization is implicitly used in~\cite{FleischerKL17tocs}. We require that $\alpha$ and $\beta$ have the same subwords of length at most $|M|$. Using the following procedure, we construct a pair of factorizations 
\begin{align*}
  \alpha &= u_1 c_1 \dots u_n c_n \alpha' 
  \\ \beta &= v_1 c_1 \dots v_n c_n \beta'
\end{align*}
Let it be assumed that we already have constructed the factorizations $\alpha = u_1 c_1 \dots u_i c_i \alpha'$ and $\beta = v_1 c_1 \dots v_i c_i \beta'$ (at the beginning, we would have $i=0$). If $u_1 c_1 \dots u_i c_i \Req u_1 c_1 \dots u_i c_i p$ for every prefix $p$ of $\beta'$, we put $n = i$ and are done. Note that we indeed consider prefixes of $\beta'$ rather than $\alpha'$. Otherwise, let $v_{i+1}c_{i+1}$ be a prefix of $\beta'$ such that $u_1 c_1 \dots u_i c_i \Req u_1 c_1 \dots u_i c_i v_{i+1} \Rl u_1 c_1 \dots u_i c_i v_{i+1} c_{i+1}$. Let $u_{i+1}$ be the longest prefix of $\alpha'$ such that $c_{i+1}$ is not in $\alp(u_{i+1})$. Note that $c_{i+1} \in \alp(\alpha')$ because the subword $c_1 \cdots c_{i+1}$ occurs in $\beta$ and hence in $\alpha$. Therefore, we obtain a new marker $c_{i+1}$ for both factorizations, and we continue the procedure with $\alpha = u_1 c_1 \dots u_i c_i u_{i+1} c_{i+1} \alpha''$ and $\beta = v_1 c_1 \dots v_i c_i v_{i+1} c_{i+1} \beta''$. 

By Lemma~\ref{lem:alphabetLR}, we have $u_1 c_1 \dots u_i c_i  \Rl u_1 c_1 \dots u_i c_i u_{i+1} c_{i+1}$. Thus the above construction will end, since the number of \greenR-classes is finite. In particular, $n < |M|$. Again by Lemma~\ref{lem:alphabetLR}, the resulting factorization satisfies $c_i \notin \alp(v_i)$ for all $1 \leq i \leq n$. In particular, the position of $c_i$ in the factorization of both $\alpha$ and $\beta$ is the defined by the first occurrence of the subword $c_1 \dots c_i$. Similarly $b_i \notin \alp(u_{i+1})$ for the \greenL-factorization above, making it the last occurrence of the subword $b_i \dots b_n$ in $\alpha$.
In both these factorizations, the positions $n_i$ corresponding to the $b_i$s and $c_i$s are called \emph{markers}.

We now have everything we need to prove the following lemma. Note that it is $v_i$ and not $u_i$ in property \emph{(\ref{it:propRIM})} below.

\begin{lemma}\label{lem:mainfactorizationlemmaIM}
	Let $m \geq 2$, $n \geq 0$ and let $\mu: A^* \to M \in \Mvar{m}$ be a homomorphism. Let $\alpha, \beta \in A^{\infty}$ with $\im(\alpha) = \im(\beta)$ and $\alpha \leq^{\Sigma^2}_{m,n+2|M|} \beta$. Then there exist factorizations
  \begin{alignat*}{3}
			\alpha & = \,&u_1 &a_1 \cdots \,&u_k &a_k \alpha' \\
			\beta & = \,&v_1 &a_1 \cdots \,&v_k &a_k \beta'
  \end{alignat*}
  with $a_i \in A$ and $\alp(\alpha') = \im(\alpha) = \alp(\beta')$ such that the following properties hold:
	\begin{enumerate}[(i)]
	\item\label{it:proplogicIM}  $\alpha' \leq^{\Sigma^{2}}_{m-1,n} \beta'$ and $u_i \leq^{\Sigma^2}_{m-1,n} v_i$ for all $1 \leq i \leq k$.
	\item\label{it:propRIM} $1 \Req  \mu(v_1)$ and $\mu(u_1a_1 \cdots u_{i-1} a_{i-1}) \Req  \mu(u_1a_1 \cdots u_{i-1} a_{i-1} v_i)$ for $1 \leq i \leq k$. Moreover, $\mu(u_1a_1 \cdots u_{k} a_{k}) \Req  \mu(u_1a_1 \cdots u_{k} a_{k} w)$ for all $w \in \alp(\alpha')^*$.
	\item \label{it:propLIM} For every linked pair $(t,f)$ with $\beta' \in [t][f]^{\omega}$ and for every $1 \leq i \leq k$ we have $\mu(a_i \cdots v_{k}a_{k}) t \Leq  \mu(v_{i} a_i \cdots v_{k} a_{k}) t$.
	\end{enumerate}
\end{lemma}

\begin{proof}
	Let $q_1 b_1 \dots q_{k'} b_{k'} \delta$ be the \greenL-factorization of $\beta$. Remember that $b_{k'}$ is the last finitely occurring letter of $\beta$. Since $\im(\alpha) = \im(\beta)$ and $\alpha \leq^{\Sigma^2}_{2,2} \beta$, the last finitely occurring letter of $\alpha$ is also $b_{k'}$ (if there were a different last finitely occurring letter $b$ in $\alpha$, then we could consider the $\Sigma^2_{2,2}$ formula saying that there exists a $b$-position such that all greater positions are in $\im(\alpha)$). Let $p_1 b_1 \dots p_{k'} b_{k'} \gamma$ be the unique factorization of $\alpha$ such that $b_{k'}$ is the last finitely occurring letter, and $b_{i} \notin \alp(p_{i+1})$ for $1 \leq i \leq k' - 1$.

	Let $\alpha=\tilde{p}_1 c_1 \dots \tilde{p}_{j} c_{j} \tilde\gamma$ and $\beta=\tilde{q}_1 c_1 \cdots \tilde{q}_j c_j \tilde\delta$ be the factorizations given by the \greenR-factorization of $(\alpha,\beta)$ with respect to $\beta$. Indeed, if $c_{i'}$ occurs before $b_i$ in $\alpha$, then $\alpha$ has the subword $c_1 \cdots c_{i'} b_{i} \cdots b_{k'}$ of length less than $2|M|$. By $\alpha \leq^{\Sigma^2}_{m,n+2|M|} \beta$, it is also a subword of $\beta$, thereby showing that $c_{i'}$ occurs before $b_i$ in $\beta$. Symmetrically (and since $m \geq 2$), we see that if $c_{i'}$ occurs before $b_i$ in $\beta$, then is also holds for $\alpha$. We also cannot have the situation where the position of $c_{i'}$ coincides with $b_i$ in one word but the position of $c_{i'}$ is greater than the position of $b_i$ in the other word; this is due to the subword $c_1 \cdots c_{i'-1} b_i \cdots b_{k'}$. Therefore, the markers occur in the same order in both words.

	Let $\alpha  = u_1 a_1 \cdots u_k a_k \alpha'$ and $\beta  = v_1 a_1 \cdots v_k a_k \beta'$ to be the combination of these factorizations. More precisely, each $a_i$ is either one of the $b_{i'}$ (in which case we say that $a_i$ is an $\greenL$-marker) or one of the $c_{i'}$ (in which case we say that $a_i$ is an $\greenR$-marker), or both. By construction, these factorizations satisfy \emph{(\ref{it:propRIM})} and \emph{(\ref{it:propLIM})}. 

	The following claims follow by repeatedly applying Lemmas~\ref{lem:simplerelativization} and~\ref{lem:dualrelativization}:
	\begin{enumerate}[\itshape(a)]
		\item \label{aaa:claim} If $a_i$ is an \greenR-marker, then $u_{i+1} \dots u_k a_k \alpha' \leq^{\Sigma^2}_{m,n + |M|} v_{i+1} \dots v_{k} a_{k} \beta'$.
		\item \label{bbb:claim} If $a_i$ is an \greenL-marker, then $u_1 a_1 \dots u_i \leq^{\Sigma^2}_{m,n + |M|} v_1 a_1 \dots v_i$.
		\item \label{ccc:claim} If $a_i$ is an \greenR-marker and $a_j$ is an \greenL-marker for $i < j$, then $u_{i+1} a_{i+1} \dots u_j \leq^{\Sigma^2}_{m,n + j - i} v_{i+1} a_{i+1} \dots v_j$.
	\end{enumerate}
	To show \emph{(\ref{it:proplogicIM})}, we distinguish between the types ($\greenR$-marker or $\greenL$-marker) of $a_{i-1}$ and $a_i$.
	\begin{itemize}
		\item If $a_{i-1}$ and $a_i$ are both \greenR-markers, we use Claim~\textit{(\ref{aaa:claim})} and a final application of Lemma~\ref{lem:simplerelativization}.
		\item If $a_{i-1}$ and $a_i$ are both \greenL-markers, we use Claim~\textit{(\ref{bbb:claim})} and a final application of Lemma~\ref{lem:dualrelativization}.
		\item If $a_{i-1}$ is an \greenR-marker and $a_i$ is an \greenL-marker, we use Claim~\textit{(\ref{ccc:claim})}.
		\item If $a_{i-1}$ is an \greenL-marker and $a_i$ is an $\greenR$-marker, then we proceed in two steps. First, we use Claim~\textit{(\ref{ccc:claim})} to get a factor $pa_{i-1}u_{i}a_iq$ of $\alpha$ and a factor $p'a_{i-1}v_{i}a_iq'$ of $\beta$ such that $a_i \notin \alp(pa_{i-1}u_i) \cup \alp(p'a_{i-1}v_i)$, $a_{i-1} \notin \alp(u_{i}a_{i}q) \cup \alp(v_ia_{i}q')$ and $pa_{i-1}u_{i}a_iq \leq^{\Sigma^2}_{m,n+2} p'a_{i-1}v_{i}a_iq'$. This allows us to use Lemma~\ref{lem:squeezedrelativisation} to conclude $u_{i} \leq^{\Sigma^2}_{m-1,n} v_{i}$
	\end{itemize}
	If $a_k$ is an $\greenR$-marker, then $\alpha' \leq^{\Sigma^2}_{m,n} \beta'$ by Claim~\textit{(\ref{aaa:claim})}. If $a_k$ is an $\greenL$-marker, then $a_k \not\in \alp(\alpha') = \alp(\beta')$ and, hence, $\alpha' \leq^{\Sigma^2}_{m-1,n} \beta'$ follows by Lemma~\ref{lem:dualrelativization}.
\end{proof}

\begin{lemma}\label{lem:mainfactorizationlemmaN}
	Let $n \geq 0$ and let $\mu: A^* \to M \in \Mvar{2}$ be a homomorphism. Let $\alpha, \beta \in A^{\infty}$ with $\im(\alpha) \neq \im(\beta)$ and $\alpha \leq^{\Sigma^2}_{2,n+2|M|} \beta$. Then $\im(\beta) \subsetneq \im(\alpha)$ and there exist factorizations
  \begin{alignat*}{3}
			\alpha & = \,&u_1 &a_1 \cdots \,&u_k &a_k \alpha' \\
			\beta & = \,&v_1 &a_1 \cdots \,&v_k &a_k v_{k+1} a_{k+1} \beta'
  \end{alignat*}
  with $a_i \in A$ and $\alp(a_{k+1}\beta') \subseteq \alp(\alpha') = \im(\alpha)$ such that the following properties hold:
	\begin{enumerate}[(i)]
	\item\label{aaa:mainfactorizationlemmaN} $\alpha' \leq^{\Sigma^2}_{1,n} v_{k+1}$ and $u_i \leq^{\Sigma^2}_{1,n} v_i$ for all $1 \leq i \leq k$.
	\item\label{bbb:mainfactorizationlemmaN} $1 \Req  \mu(v_1)$ and $\mu(u_1a_1 \cdots u_{i-1} a_{i-1}) \Req  \mu(u_1a_1 \cdots u_{i-1} a_{i-1} v_i)$ for $1 \leq i \leq k$. Moreover, $\mu(u_1a_1 \cdots u_{k} a_{k}) \Req  \mu(u_1a_1 \cdots u_{k} a_{k} w)$ for all $w \in \alp(\alpha')^*$.
	\item \label{ccc:mainfactorizationlemmaN} For every linked pair $(t,f)$ with $\beta' \in [t][f]^{\omega}$ and for every $1 \leq i \leq k+1$ we have $\mu(a_i \cdots v_{k+1} a_{k+1}) t \Leq  \mu(v_{i} a_i \cdots v_{k+1} a_{k+1}) t$.
	\end{enumerate}
\end{lemma}

\begin{proof}
	It is easy to see that $\alp(\alpha) = \alp(\beta)$. If $a \in \alp(\alpha)$ occurs only finitely often in $\alpha$, then $\alpha$ satisfies the formula
	\begin{equation*}
		\exists x : \lambda(x) = a \wedge \neg \exists y : x < y \wedge \lambda(y) = a 
	\end{equation*}
	which says that there exists a last occurrence of the letter $a$. Since $\alpha \leq^{\Sigma^2}_{2,n+2|M|} \beta$, the word $\beta$ also satisfies the above formula, thereby showing that $a$ occurs only finitely often in $\beta$. This shows that $\alp(\alpha) \setminus \im(\alpha)$ is a subset of $\alp(\beta) \setminus \im(\beta)$. Hence, $\im(\alpha) \neq \im(\beta)$ yields $\im(\beta) \subsetneq \im(\alpha)$.

	Let $a$ be the last finitely occurring letter of $\alpha$ (if it exists). Let $ \beta = q_1b_1 \dots q_{\ell}b_{\ell}\delta$ be the \greenL-factorization of $\beta$, but if the last occurrence of $a$ in $\beta$ is not a marker of the $\greenL$-factorization, then we additionally include this last occurrence of $a$ as a marker in the factorization. Let $b_{k'}$ be the last occurrence of $a$ in $\beta$ (if there is no last finitely occurring letter $a$ in $\alpha$, then we can imagine that $k'=0$ and $b_0 = \varepsilon$). Let $\alpha = p_1 b_1 \dots p_{k'} b_{k'} \gamma$ be the unique factorization of $\alpha$ such that $b_{k'} \notin \alp(\gamma)$ and $b_{i} \notin \alp(p_{i+1})$ for $1 \leq i < k'$. Let $b \in \im(\alpha)$ and $c \notin \im(\alpha)$. Then $\alpha$, and hence $\beta$, satisfies the following formulae
	\begin{align*}
		\varphi_b & = \exists x : \lambda(x) = b_{k'} \wedge (\forall y : y \leq x \vee \neg \lambda(y) = b_{k'}) \wedge (\exists y  : x < y \wedge \lambda(y) = b)
		\\ \varphi_c & = \exists x : \lambda(x) = b_{k'} \wedge (\forall y : y \leq x \vee \neg \lambda(y) = b_{k'}) \wedge (\forall y  : y \leq x \vee \neg \lambda(y) = c).
	\end{align*}
	This shows that $\alp(p_{k'+1} b_{k'+1} \cdots p_\ell b_\ell \delta) = \im(\alpha)$. In particular, we have $k' < \ell$ (i.e., $p_{k'+1}b_{k'+1} \dots p_{\ell} b_{\ell}$ is nonempty) because $\alp(\delta) = \im(\beta) \subsetneq \im(\alpha)$. Furthermore, there are infinitely many occurrences of the letters $b_{k'+1}, \ldots, b_{\ell}$ in $\gamma$.
	
	 Let $\alpha=\tilde{p}_1 c_1 \dots \tilde{p}_{j} c_{j} \tilde\gamma$ and $\beta=\tilde{q}_1 c_1 \cdots \tilde{q}_j c_j \tilde\delta$ be the factorizations given by the \greenR-factorization of $(\alpha,\beta)$ with respect to $\beta$. 
	We can use the same argument as in Lemma~\ref{lem:mainfactorizationlemmaIM} to conclude that the markers $b_i$ for $1 \leq i \leq k'$ and the markers $c_i$ for $1 \leq i \leq j$ occur in the same order in both $\alpha$ and $\beta$. Moreover, the marker $b_{k'+1}$ occurs to the right of $c_j$ in $\beta$ because $c_1 \cdots c_j b_{k'+1} \cdots b_{\ell}$ is a subword of $\alpha$ and, hence, of $\beta$. Let $\alpha  = u_1 a_1 \cdots u_k a_k \alpha'$ be the factorization of $\alpha$ at all markers $b_i$ for $1 \leq i \leq k'$ and all markers $c_i$ for $1 \leq i \leq j$; these markers define the letters $a_i$. Similarly, let $\beta  = v_1 a_1 \cdots v_k a_k v_{k+1} a_{k+1} \beta'$ be the factorization of $\beta$ at all markers $b_i$ for $1 \leq i \leq k'+1$ and all markers $c_i$ for $1 \leq i \leq j$; since $b_{k'+1}$ occurs after $c_j$, we have $a_{k+1} = b_{k'+1}$. These factorizations satisfy properties \emph{(\ref{bbb:mainfactorizationlemmaN})} and \emph{(\ref{ccc:mainfactorizationlemmaN})} by construction.
	
	By choice of $b_{k'}$, we have $\alp(\alpha') = \im(\alpha') = \im(\alpha)$ and $\alp(a_{k+1}\beta') \subseteq \alp(\alpha')$. Repeatedly applying Lemma \ref{lem:dualrelativization2} yields
	\begin{equation}\label{eqn:mainfactorizationlemmaN}
		u_1 a_1 \dots u_k a_k \alpha' \;\leq^{\Sigma^2}_{m,n + 2|M| + k' - \ell}\; v_1 a_1 \dots v_k a_k v_{k+1}\,.
	\end{equation}
	Since the rightmost \greenL-marker of these two words is the last occurrence of this letter in both words, we get invariants similar to \textit{(\ref{aaa:claim})}, \textit{(\ref{bbb:claim})} and \textit{(\ref{ccc:claim})} from the proof of Lemma~\ref{lem:mainfactorizationlemmaIM}. Thus we get $u_i \leq^{\Sigma^2}_{1,n} v_i$ for all $1 \leq i \leq k$ by the same argument as in the proof of that lemma.
	Finally, we get $\alpha \leq^{\Sigma^2}_{1,n} v_{k+1}$ by applying Lemma \ref{lem:dualrelativization} to the words in Equation~(\ref{eqn:mainfactorizationlemmaN}) using the \greenL-marker $b_{k'}$, possibly followed by multiple applications of Lemma~\ref{lem:simplerelativization} for eliminating the \greenR-markers to the right of $b_{k'}$.
\end{proof}

We  prove that for every language $L$ recognized by a monoid in $\Mvar{m}$, there exists $n$ satisfying the following implication: if $\alpha \leq_{m,n}^{\Sigma^2} \beta$ for $\im(\alpha) = \im(\beta)$ and if $\alpha \in L$, then $\beta \in L$. Since the infinite alphabets always agree when $m \geq 3$, this is enough to prove the remaining parts of Theorem~\ref{thm:sigmatwo} in these cases. The case $m = 2$ and $\im(\alpha) \neq \im(\beta)$ is more involved.

\begin{lemma}\label{lem:maininductionstep}
	Let $\mu: A^* \to M \in \Mvar{m}$ for $m \geq 1$. Then there exists $n$ such that $\leq^{\Sigma^2}_{m,n}$ has the following property: If two words $\alpha,\beta \in A^\infty$ with $\im(\alpha) = \im(\beta)$ satisfy $\alpha \leq^{\Sigma^2}_{m,n} \beta$, then all linked pairs $(s,e)$, $(t,f)$ in $M$ with $\alpha \in [s][e]^{\omega}$ and $\beta \in [t][f]^{\omega}$ satisfy $(s,e) \lesssim (t,f)$.
\end{lemma}

\begin{proof}
	We proceed by induction on $m$. Consider the base case $m=1$. By Theorem~\ref{thm:sigmatwo}\,\textit{(\ref{aaa:sigmatwo})}, there exists $n$ such that every open language recognized by $\mu$ is defined by a formula in $\Sigma^2_1$ of depth at most $n$. Consider $\alpha,\beta \in A^\infty$ with $\im(\alpha) = \im(\beta)$ and $\alpha \leq^{\Sigma^2}_{1,n} \beta$. Suppose that $(s,e)$ and $(t,f)$ are linked pairs in $M$ with $\alpha \in [s][e]^{\omega}$ and $\beta \in [t][f]^{\omega}$. Since $\Sigma^2_1$-definable languages are open in the Cantor topology, we can choose a sufficiently long prefix $u$ of $\alpha$ such that the factorization $\alpha = u \alpha'$ satisfies
	\begin{itemize}
	\item $u \in A^*$ satisfies the same sentences in $\Sigma^2_{1,n}$ as $\alpha$,
	\item $\im(\alpha) = \alp(\alpha')$, and
	\item $u \in [s]$ and $\alpha' \in [e]^\omega$
	\end{itemize}
	Similarly, let $\beta = v \beta'$ such that
	\begin{itemize}
	\item $v \in A^*$ satisfies the same sentences in $\Sigma^2_{1,n}$ as $\beta$,
	\item $\im(\beta) = \alp(\beta')$, and
	\item $v \in [t]$ and $\beta' \in [f]^\omega$
	\end{itemize}
	In particular, $u \leq^{\Sigma^2}_{1,n} v$.
	Consider the language $L = \bigcup_{s\leq r} [r] A^\infty$; it is open and recognized by $M$. Now, $u \in L$ yields $v \in L$. Thus, there exists $r \in M$ with $s \leq r$ and $v \in [r]A^*$. Hence, there exists $z \in M$ with $t=rz$. We conclude that $s \leq r \leq rz = t$. The inequality $r \leq rz$ follows by $M \in \Mvar{1} = \intp{1 \leq z}$. In addition, $e \leq fef$ and $f \leq efe$ by $M \in \Mvar{1}$. Since $\im(\alpha) = \im(\beta)$, Lemma~\ref{lem:DA} yields $fef = f$ and $efe = e$. Thus $e=f$. This shows that $(s,e) = (s,f) \leq (t,f)$ and in particular $(s,e) \lesssim (t,f)$.
	
	Let now $m \geq 2$. Let $\pi: M \to \faktor{M}{\preceq_{\mathbf{KD}}}$ be the canonical projection and let $\mu' = \pi \circ \mu$. Note that $\mu' : A^* \to \faktor{M}{\preceq_{\mathbf{KD}}} \in \Mvar{m-1}$. By induction, there exists an integer $n'$ satisfying the claim of the lemma for $\mu'$. Let $n = n' + 2|M|$, and let $\alpha$ and $\beta$ be as in the statement of the lemma. Lemma~\ref{lem:mainfactorizationlemmaIM} yields factorizations $\alpha = u_1a_1 \cdots u_{k}a_{k} \alpha'$ and $\beta = v_1a_1 \cdots v_{k} a_{k} \beta'$.  Let $u = u_1a_1 \dots u_{k} a_{k}$, $v = v_1a_1 \dots v_{k} a_{k}$, and let $\alpha' \in [s'][e]^{\omega}$ and $\beta' \in [t'][f]^{\omega}$ for linked pairs $(s',e)$, $(t',f)$ with $s=\mu(u)\hspace*{1pt}s'$ and $t=\mu(v)\hspace*{1pt}t'$. 

	We have $\im(\alpha') = \im(\alpha) = \im(\beta) = \im(\beta')$ and $\alpha' \leq^{\Sigma^2}_{m-1,n'} \beta'$ by Lemma~\ref{lem:mainfactorizationlemmaIM}. By choice of $n'$, there exists $r' \in M$ such that $(\pi(s'),\pi(e)) \sim (\pi(r'),\pi(f)) \leq (\pi(t'),\pi(f))$; in particular, we have $r' \preceq_{\mathbf{KD}} t'$.  Since $\alp(\beta') = \im(\beta')$, there exist representatives $\hat{t} \in [t']$, $\hat{f} \in [f]$ such that $\alp(\hat{t}) \subseteq \alp(\hat{f})$. We thus have $ft'f = f$ by Lemma \ref{lem:DA}. In particular, we have $t'f \Leq f$. By Lemma~\ref{lem:mainfactorizationlemmaIM}, we have $\mu(u) \hspace*{1pt} t' \Req \mu(u)$. It follows that $\mu(u) \hspace*{1pt} r' f \leq \mu(u) \hspace*{1pt} t' f = \mu(u) \hspace*{1pt} t'$. 
	
	We want to show that $(s,e) \sim (r,f) \leq (t,f)$ for $r = \mu(u)\hspace*{1pt} r' f$. This consists of two parts: We first prove that $(s,e) \sim (r,f)$, and then we verify that $\mu(u) \hspace*{1pt} t' \leq \mu(v) \hspace*{1pt} t'$. Since $r = \mu(u) \hspace*{1pt} r' f \leq \mu(u) \hspace*{1pt} t'$ and $\mu(v) \hspace*{1pt} t' = t$, the second part then yields $r \leq t$. 
	
	Since $(\pi(s'),\pi(e)) \sim (\pi(r'),\pi(f))$, there exist $x,y \in M$ such that $\pi(e) = \pi(xy)$, $\pi(f) = \pi(yx)$ and $\pi(s'x) = \pi(r')$, i.e., $s'x \equiv_{\mathbf{KD}} r'$. Let $p = exf$ and $q = fye$. Since $r' \preceq_{\mathbf{KD}} t'$,  we have $\mu(u)r' \Req \mu(u)$ and $rf \Leq f$. Now, Lemma~\ref{lem:liftingEQ} yields $sp = \mu(u)s'exf = \mu(u)s'xf = \mu(u)r'f=r$. Lemma~\ref{lem:DA} yields $e = pq$ and $f = qp$ because $\alp(\alpha') = \im(\alpha') = \im(\alpha) = \im(\beta) = \im(\beta') = \alp(\beta')$ by Lemma~\ref{lem:mainfactorizationlemmaIM}. Therefore, the linked pairs $(s,e)$ and $(r,f)$ are conjugated via~$p,q$.
	
	It remains to show that $\mu(u) \hspace*{1pt} t' \leq \mu(v) \hspace*{1pt} t'$. Induction together with Lemma~\ref{lem:makefinite} yields $\mu(u_i) \preceq_{\mathbf{KD}} \mu(v_i)$ for all $1 \leq i \leq k$. Thus, the $\greenR$-equivalences and the $\greenL$-equivalences in Lemma~\ref{lem:mainfactorizationlemmaIM} allow us to successively replace $u_i$ by $v_i$ from right to left:
	\begin{align*}
			\mu(u) \hspace*{1pt} t' =
			\mu(u_1a_1 \cdots u_{k}a_{k}) \hspace*{1pt} t' & \leq
			\mu(u_1a_1 \cdots u_{k-1} a_{k-1} v_{k}a_{k}) \hspace*{1pt} t'
			\\ & \leq \mu(u_1a_1 \cdots v_{k-1} a_{k-1} v_{k}a_{k}) \hspace*{1pt} t'
			\\ & \;\; \vdots
			\\ & \leq \mu(v_1a_1 \cdots v_{k}a_{k}) \hspace*{1pt} t'
			= \mu(v) \hspace*{1pt} t'
	\end{align*}
	This shows $(r,f) \leq (t,f)$ and, hence, $(s,e) \lesssim (t,f)$.
\end{proof}

For $m=2$ and if $\im(\alpha) \neq \im(\beta)$, we have the following slightly weaker version of Lemma~\ref{lem:maininductionstep}. This weaker property is sufficient for proving Theorem~\ref{thm:sigmatwo}, but we needed a stronger invariant for the induction step in Lemma~\ref{lem:maininductionstep}.

\begin{lemma}\label{lem:leveltwo}
  Let $L \subseteq A^\infty$ be open in the alphabetic topology and recognized by $\mu: A^* \to M \in \Mvar{2}$. Then there exists an integer $n$ such that whenever $\alpha \leq^{\Sigma^2}_{2,n} \beta$ for $\alpha,\beta \in A^\infty$, then $\alpha \in L$ implies $\beta \in L$.
\end{lemma}

\begin{proof}
  By Corollary~\ref{cor:sigmatwostar}, there exists $n'$ such that $u \leq^{\Sigma^2}_{1,n'} v$ implies $\mu(u) \preceq_{\mathbf{KD}} \mu(v)$ for all finite words $u,v \in A^*$. Let $n$ be the maximum of $n' + 2 \left| M \right|$ and the number given by Lemma~\ref{lem:maininductionstep} and suppose that $\alpha \leq^{\Sigma^2}_{2,n} \beta$. If $\im(\beta) = \im(\alpha)$, then the result follows by Lemma~\ref{lem:maininductionstep}. It therefore suffices to consider the case $\im(\beta) \neq \im(\alpha)$. Then $\im(\beta) \subsetneq \im(\alpha)$ by Lemma~\ref{lem:mainfactorizationlemmaN}. In particular, $\alpha$ is infinite. Applying Lemma~\ref{lem:mainfactorizationlemmaN}, yields factorizations $\alpha = u_1a_1 \cdots u_k a_k \alpha'$, $\beta = v_1a_1 \cdots v_k a_k v_{k+1} a_{k+1} \beta'$. Let $u = u_1a_1 \cdots u_k a_k$ and $v = v_1a_1 \cdots v_k a_k v_{k+1}$.
	
	Since $\alpha$ is infinite, it has arbitrarily long prefixes. Thus, we can choose $\hat{s} \in A^*$ such that $\alpha' = \hat{s} \alpha''$ and the following properties are satisfied:
	\begin{enumerate}[\itshape(i)]
	\item\label{it:sconditionone} $u\hat{s}\gamma \in L$ for all $\gamma \in \im(\alpha)^{\infty}$ (since $L$ is open in the alphabetic topology),
	\item\label{it:sconditiontwo} $\hat{s}$ contains the same subwords of length $n'$ as $\alpha'$.
	\end{enumerate}
	By Lemma \ref{lem:mainfactorizationlemmaN}, we have $\alp(a_{k+1}\beta') \subseteq \im(\alpha)$ and $\alpha \leq^{\Sigma^2}_{1,n'} v_{k+1}$. The former yields $u \hat{s} a_{k+1}\beta' \in L$ by property \textit{(\ref{it:sconditionone})}, and the latter yields $\mu(\hat{s}) \preceq_{\mathbf{KD}} \mu(v_{k+1})$ by property \textit{(\ref{it:sconditiontwo})}. Let $a_{k+1}\beta' \in [t][f]^\omega$ for some linked pair $(t,f)$. By Lemma~\ref{lem:mainfactorizationlemmaN}, we have $\mu(u v_{k+1}) \Req \mu(u)$ and $\mu(v_{k+1}) t \Leq t$, and thus
	\begin{equation*}
		\mu(u \hat{s})t \leq \mu(u v_{k+1}) t
	\end{equation*}
	By Lemma~\ref{lem:mainfactorizationlemmaN}, we also have $u_{i} \leq^{\Sigma^2}_{1,n'} v_{i}$ for all $1 \leq i \leq k$. Thus, $\mu(u_i) \preceq_{\mathbf{KD}} \mu(v_i)$ for all $1 \leq i \leq k$. By repeatedly applying the definition of $\preceq_{\mathbf{KD}}$, we obtain
	\begin{equation*}
		\begin{aligned}
		\mu(u v_{k+1}) t \,&= \mu(u_1a_1 \cdots u_{k-1} a_{k-1} u_k a_kv_{k+1})t 
		\\ & \leq \mu(u_1a_1 \cdots u_{k-1} a_{k-1} v_k a_kv_{k+1})t
		\\ & \leq \mu(u_1a_1 \cdots v_{k-1} a_{k-1} v_k a_kv_{k+1})t
		\\ & \;\; \vdots 
		\\ & \leq \mu(v_1a_1 \cdots v_k a_kv_{k+1})t = \mu(v)t
		\end{aligned}
	\end{equation*}
	and, hence, $\mu(u \hat{s})t \leq \mu(v)t$.
	Putting everything together, we get
  \begin{alignat*}{2}
	  \alpha = u\hat{s}\alpha'' \in L
	  \quad \Rightarrow \quad & u\hat{s}a_{k+1}\beta \in L
    &\qquad\quad&\text{by property \textit{(\ref{it:sconditionone})}}
    \\ \Rightarrow \quad &  [\mu(u\hat{s})t][f]^{\omega} \subseteq L
    &&\text{by recognition}
    \\ \Rightarrow \quad & [\mu(v) t][f]^{\omega} \subseteq L
    &&\text{by upward closure}
    \\ \Rightarrow \quad & \beta \in L
    &&\text{since $\beta = v \beta' \in [\mu(v) t][f]^{\omega}$}
  \end{alignat*}
  which is the desired result.
\end{proof}

We are now ready to conclude the main results.

\begin{proof}[Proof of Theorem~\ref{thm:sigmatwo} (\ref{bbb:sigmatwo}) and (\ref{ccc:sigmatwo})]
  Assume that $L$ is definable in $\Sigma^2_m$. Proposition~\ref{prp:logic2algebra} shows that $M \in \Mvar{2}$. If $L$ is definable in $\Sigma^2_{2}$, then $L$ is definable in $\Sigma_2$ and, hence, $L$ is open in the alphabetic topology by Lemma~\ref{lem:alwaysopeninthealphabetictopology}. This concludes the implications from left to right in both \emph{(\ref{bbb:sigmatwo})} and \emph{(\ref{ccc:sigmatwo})}.

	For the other direction, let $m \geq 2$ and let $\mu: A^* \to M \in \Mvar{m}$ recognize $L \subseteq A^{\infty}$. Moreover, in the case of $m=2$ we assume that $L$ is open in the alphabetic topology. Let $n$ be the integer given by Lemma~\ref{lem:maininductionstep} or Lemma~\ref{lem:leveltwo}, respectively. Note that if $m \geq 3$, then $\alpha \leq^{\Sigma_2}_{m,n} \beta$ implies $\im(\alpha) = \im(\beta)$. Therefore, whenever $\alpha \leq^{\Sigma^2}_{m,n} \beta$, then $\alpha \in L$ implies $\beta \in L$. Let $\varphi_{\alpha}$ denote the conjunction of all formulae in $\Sigma^2_{m,n}$ which are true on $\alpha$. Since the number of inequivalent formulae in $\Sigma^2_{m,n}$ is finite (see e.g.~\cite[Prop.~IV.1.1]{str94}), this conjunction is a well-defined $\Sigma^2_{m,n}$-formula. Consider the disjunction $\varphi = \bigvee_{\alpha \in L} \varphi_{\alpha}$. Again, this disjunction is finite. We show that $\varphi$ defines $L$. The inclusion $L \subseteq L(\varphi)$ hold by definition of $\varphi$. Suppose that $\beta \models \varphi$. Then there exists $\alpha \in L$ with $\beta \models \varphi_{\alpha}$. By construction of $\varphi_\alpha$ we have $\alpha \leq^{\Sigma^2}_{m,n} \beta$ and, hence, $\beta \in L$.
\end{proof}

\section{Conclusion}

We introduced the preorder $\preceq_{\mathbf{KD}}$ and used it for constructing an infinite hierarchy $\Mvar{m}$ of finite ordered monoids inside $\DA$. This hierarchy was then used for characterizing all half-levels $\Sigma^2_m$ of the quantifier alternation hierarchy inside two-variable first-order logic $\FO^2$ over $A^\infty$, the class of all finite and infinite words over the alphabet $A$. Our contribution uses novel techniques and generalizes the corresponding results for finite words~\cite{FleischerKL17tocs}.

In Proposition~\ref{prp:intersectionsoftwh}, we showed the equivalence between $\Mvar{m}$ and the previously introduced varieties $\intp{V_m \leq U_m}$. The latter characterization has the advantage of giving a very direct insight of the structure of the languages definable in each logical fragment. One possible application could be in the analysis of Carton-Michel automata, also known as complete unambiguous automata or prophetic automata. In \cite{preugschatwilke2013lmcs}, Preugschat and Wilke use forbidden patterns together with certain conditions on so called \emph{loop languages} to characterize fragments of temporal logic. We conjecture that it is possible to find forbidden patterns for the languages $\Sigma^2_m$ by leveraging knowledge of the algebraic structure, and conditions on the loop languages using the topological properties. In particular, this could give a more efficient decision procedure, without having to rely on the (possibly exponentially bigger) transition monoid.

\smallskip


\end{document}